\documentclass[prd,aps,twocolumn,superscriptaddress,showpacs,preprintnumbers,amsmath,amssymb,floatfix,notitlepage]{revtex4-1}
\usepackage{graphicx} 

\begin{document}
%%%%%%%%%%%%%%%%%%%%%%%%%%%%%%%%%%%%%%%%%%%%%%%%%%%%%%%%%%%%%%%%%%%%%%%%%%%%%%%%%%%%%%%
\title{A New Method for Measuring Coherent Elastic Neutrino Nucleus Scattering\\ at an Off-Axis High-Energy Neutrino Beam Target}	

%%%%%%%%%%%%%%%%%%%%%%%%%%%%%%%%%%%%%%%%%%%%%%%%%%%%%%%%%%%%%%%%%%%%%%%%%%%%%%%%%%%%%%%
\author{S.J.~Brice}    \affiliation{Fermi National Accelerator Laboratory, Batavia, IL, 60510, USA}
\author{R.L.~Cooper}   \affiliation{Indiana University, Bloomington, IN, 47405, USA}
\author{F.~DeJongh}    \affiliation{Fermi National Accelerator Laboratory, Batavia, IL, 60510, USA}
\author{A.~Empl}       \affiliation{University of Houston, Houston, TX, 77204, USA}
\author{L.M.~Garrison} \affiliation{Indiana University, Bloomington, IN, 47405, USA}
\author{A.~Hime}       \affiliation{Los Alamos National Laboratory, Los Alamos, NM 87545, USA}
\author{E.~Hungerford} \affiliation{University of Houston, Houston, TX, 77204, USA}
\author{T.~Kobilarcik} \affiliation{Fermi National Accelerator Laboratory, Batavia, IL, 60510, USA}
\author{B.~Loer}       \affiliation{Fermi National Accelerator Laboratory, Batavia, IL, 60510, USA}
\author{C.~Mariani}    \affiliation{Virginia Tech, Blacksburg, VA 24061, USA}
\author{M.~Mocko}      \affiliation{Los Alamos National Laboratory, Los Alamos, NM 87545, USA}
\author{G.~Muhrer}     \affiliation{Los Alamos National Laboratory, Los Alamos, NM 87545, USA}
\author{R.~Pattie}     \affiliation{North Carolina State University, NC 27695, USA}
\author{Z.~Pavlovic}   \affiliation{Los Alamos National Laboratory, Los Alamos, NM 87545, USA}
\author{E.~Ramberg}    \affiliation{Fermi National Accelerator Laboratory, Batavia, IL, 60510, USA}
\author{K.~Scholberg}  \affiliation{Duke University, Durham, NC, 27708, USA}
\author{R.~Tayloe}     \affiliation{Indiana University, Bloomington, IN, 47405, USA}
\author{R.T.~Thornton} \affiliation{Indiana University, Bloomington, IN, 47405, USA}
\author{J.~Yoo}        \affiliation{Fermi National Accelerator Laboratory, Batavia, IL, 60510, USA}
\author{A.~Young}      \affiliation{North Carolina State University, NC 27695, USA}
%%%%%%%%%%%%%%%%%%%%%%%%%%%%%%%%%%%%%%%%%%%%%%%%%%%%%%%%%%%%%%%%%%%%%%%%%%%%%%%%%%%%%%%

\begin{abstract}
	We present a new experimental method for measuring the process of Coherent Elastic Neutrino Nucleus Scattering (CENNS). This method uses a detector situated transverse to a high energy neutrino beam production target. This detector would be sensitive to the low energy neutrinos arising from pion decays-at-rest in the target. We discuss the physics motivation for making this measurement and outline the predicted backgrounds and sensitivities using this approach. We report a measurement of neutron backgrounds as found in an off-axis surface location of the Fermilab Booster Neutrino Beam (BNB) target. The results indicate that the Fermilab BNB target is a favorable location for a CENNS experiment. \par
\end{abstract}

\pacs{29.25.-t, 13.15.+g, 13.40.Em, 23.40.Bw, 95.35.+d}
\preprint{FERMILAB-PUB-13-522-E}
\maketitle

%%%%%%%%%%%%%%%%%%%%%%%%%%%%%%%%%%%%%%%%%%%%%%%%%%%%%%%%%%%%%%%%%%%%%%%%%%%%%%%%%%%%%%%
\section{Introduction}\label{intro}
%%%%%%%%%%%%%%%%%%%%%%%%%%%%%%%%%%%%%%%%%%%%%%%%%%%%%%%%%%%%%%%%%%%%%%%%%%%%%%%%%%%%%%%
The Coherent Elastic Neutrino-Nucleus Scattering process, or CENNS, has yet to be observed since its first theoretical prediction in 1974 by D.~Freedman~\cite{Freedman:1973yd}. The condition of coherence requires sufficiently small momentum transfer to a nucleon so that the waves of off-scattered nucleons in the nucleus are all in phase and add up coherently. Neutrinos with energies less than 50\,MeV largely fulfill this coherence condition in most target materials. The elastic neutral current interaction leaves no observable signature other than the low-energy recoils of the nucleus with energies of up to tens of keV. The technical difficulties of developing large-scale, low-energy threshold, and low-background detectors have hampered the experimental realization of the CENNS measurement for more than four decades. However, recent innovations in dark matter detector technology have made the unseen CENNS testable. \par

Neutrinos and dark matter are similar in that they both exist ubiquitously in the Universe and interact very weakly. All major dark matter direct detection searches rely on the postulation of coherent scattering of these massive particles off of nuclei. Because of the relatively low momentum transfer, the total interaction cross-section scales as the atomic mass squared of the target material. This is an analogy for low-energy neutrinos interacting coherently with nuclei. In fact, the CENNS interactions may prove to be an irreducible background for future direct detection dark matter searches. \par

Besides its role as a fundamental background in dark matter searches, measurement of the CENNS process impacts a significant number of physics and astrophysics topics, including supernova explosions, neutron form factor, sterile neutrino, neutrino magnetic moments and other non-Standard Model physics. \par

The method we outline uses low energy neutrinos arising from pion decay-at-rest source in the existing high energy neutrino beam~\cite{Yoo:2011}. This differs from other methods for which detectors are proposed to be situated close to the core of a nuclear reactor~\cite{Wong:2005vg,Barbeau:2007qi} or spallation neutron sources~\cite{Scholberg:2009ha,Akimov:2013yow}. \par

In this paper, we present R\&D for a measurement of CENNS. We start by discussing the physics motivation for the CENNS process in section~\ref{physics}. The details of the high-intensity and low-energy neutrino flux from the Fermilab Booster Neutrino Beam (BNB) is explained in section~\ref{bnb}. The beam-associated background and cosmogenic background measurements at the BNB target building are described in section~\ref{neutronbg}, a conceptual CENNS experiment is described in section~\ref{cennsexperiment}, and we summarize this paper in section~\ref{summary}.

%%%%%%%%%%%%%%%%%%%%%%%%%%%%%%%%%%%%%%%%%%%%%%%%%%%%%%%%%%%%%%%%%%%%%%%%%%%%%%%%%%%%%%%
\section{Physics Motivation}\label{physics}
%%%%%%%%%%%%%%%%%%%%%%%%%%%%%%%%%%%%%%%%%%%%%%%%%%%%%%%%%%%%%%%%%%%%%%%%%%%%%%%%%%%%%%%
%FIG==============
\begin{figure}[t!]
\centering
\includegraphics[width=1.5in]{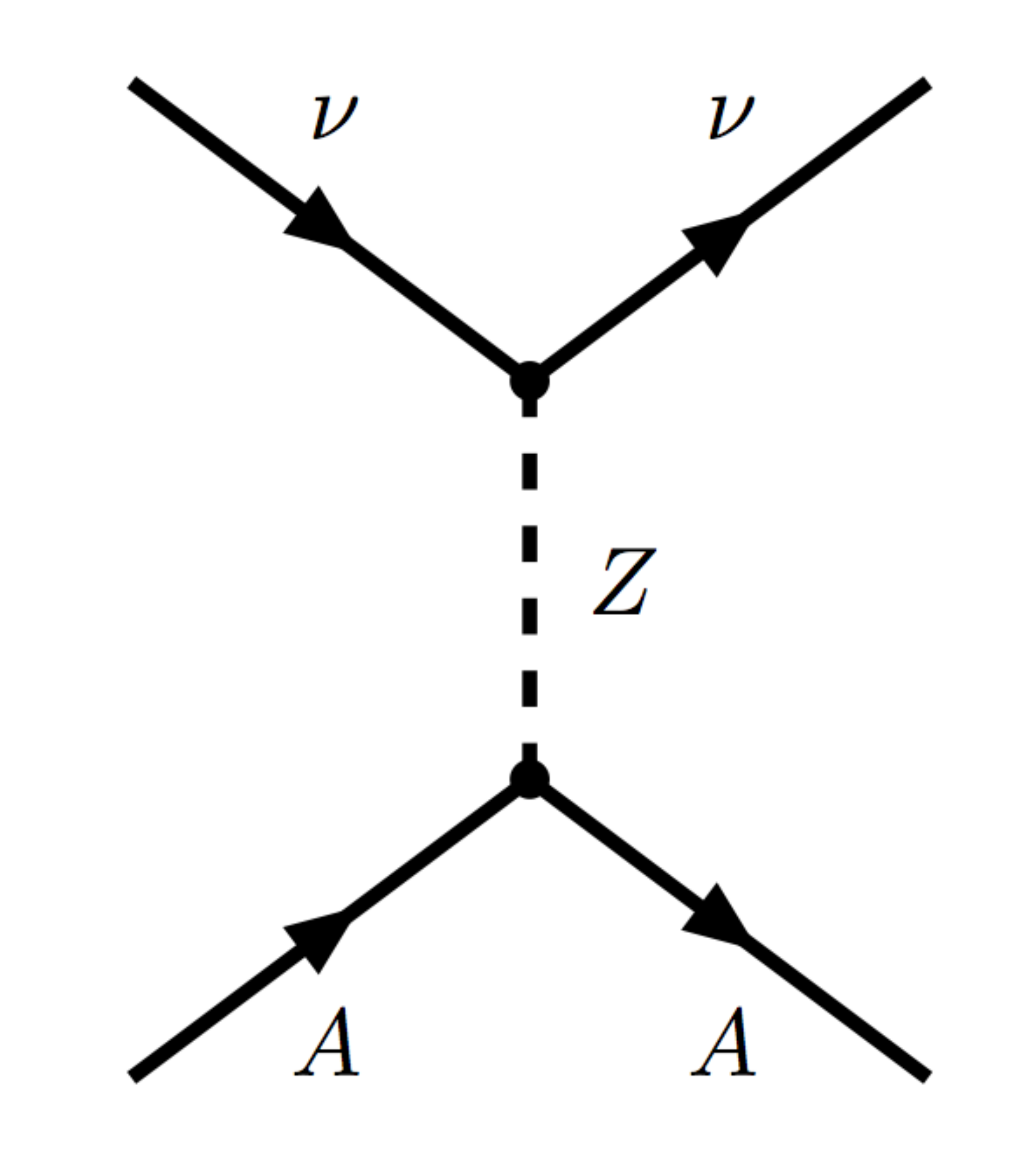}
{\caption{Feynman diagram of the CENNS process.}\label{fig:cenns}}
\end{figure}
%FIG==============
	 In the Standard Model, CENNS is mediated by $Z^0$ vector boson exchange (see FIG.~\ref{fig:cenns}). In this process a neutrino of any flavor scatters off a nucleus with the same strength; hence, the measurement will be insensitive to neutrino flavor and will be blind to neutrino oscillations among the active flavors. The dominant cross section for a spin-zero nucleus at an incident neutrino energy of $E_\nu$ is given by
\begin{equation}
\sigma_{\nu A} \simeq \frac{4}{\pi} E_\nu ^2 [Z w_p + (A-Z) w_n]^2,
	\label{eqn:sigmanuN}
\end{equation}
\noindent where the $Z$ is an atomic number and $A$ is an atomic mass. $\nu A$ stands for neutrino-nuclei interaction. The vector charge of $Z^0$ to $u$-quark ($\frac{1}{4}-\frac{2}{3}\sin^2\theta_{\mbox{w}}$) and $Z^0$ to $d$-quark ($-\frac{1}{4}+\frac{1}{3}\sin^2\theta_{\mbox{w}}$), where $\theta_{\mbox{w}}$ is the Weinberg angle, causes the different coupling strength between $w_p$ and $w_n$ to the proton ($uud$) and the neutron ($udd$), respectively. The SM values are $w_p = \frac{G_F}{4} (4\sin^2 \theta_{\mbox{w}} -1)$ and $w_n = \frac{G_F}{4}$. Since $\sin^2 \theta_{\mbox{w}} \simeq 0.23$, $w_p$ is suppressed and the $\nu A$ cross section at a given neutrino energy is effectively proportional to the square of the number of neutrons, $(A-Z)^2$.\par
\begin{figure}[t]
\centering
\includegraphics[width=3.5in]{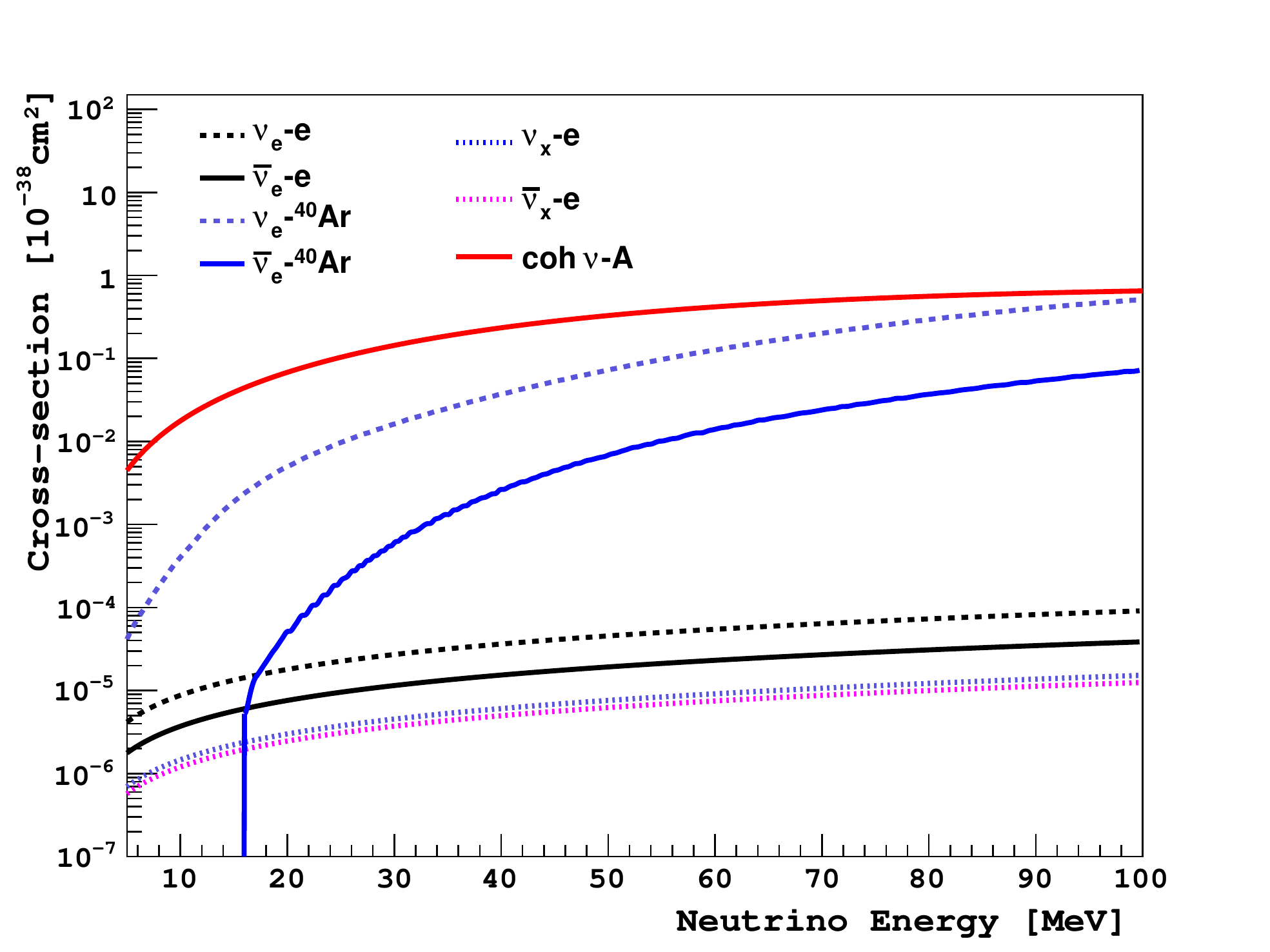}
{\caption{ Neutrino cross sections on argon target in low energy region.}\label{fig:ncrosssection}}
\end{figure}

Typical values of the total CENNS cross section for medium A nuclei are in the range of $\sim$$10^{-39}$\,cm$^2$ which is at least an order of magnitude larger than other neutrino interactions in this energy range (see FIG.~\ref{fig:ncrosssection}). For example, charged current inverse $\beta$ decay on protons has a total cross section of $\sigma_{\bar{\nu}_e p} \simeq 10^{-40}$\,cm$^2$ and elastic neutrino-electron scattering has a total cross section of $\sigma_{\nu_e e}\simeq 10^{-43}$\,cm$^2$. The maximum nuclear recoil energy for a target nucleus of mass $M$ is given by $2 E_{\nu}^2/M$ which is in the sub-MeV range for $E_{\nu}$$\sim$50\,MeV and for typical detector materials. \par
In the following sub-sections we briefly summarize the important physics cases where the CENNS interactions play a significant role.

%FIG==============
%FIG==============

%=====================================================================================
\subsection{CENNS in Particle Astrophysics}
%=====================================================================================
%-------------------------------------------------------------------------------------
\subsubsection{Dark Matter Physics}
%-------------------------------------------------------------------------------------
	One of the most fascinating problems in Particle Astrophysics is the presence of dark matter. The Standard Model (SM) does not accommodate a suitable dark matter particle candidate; therefore dark matter is crucial phenomenological evidence for physics Beyond the Standard Model (BSM). The common theme of BSM scenarios is the introduction of new particles where at least one is neutral and stable. These new particles in most scenarios typically have non-gravitational interactions which are sufficient to keep them in thermal equilibrium in the early universe. In particular, particles with a mass of the electroweak scale have a relic density in the right range for a suitable candidate for dark matter. \par

In the limit of vanishing momentum transfer the dark matter to nuclei ($\chi$$A$) cross section becomes
\begin{eqnarray}
	\sigma_{\chi A} \simeq \frac{4}{\pi} \mu^2_{\chi A} \left[Z f_p + (A-Z)f_n\right]^2 ,
	\label{eqn:sigmachiN}
\end{eqnarray}

\noindent where $\mu_{\chi A}$ is the reduced mass of the collision. A spin-independent $\chi$$A$ interaction corresponds to a coupling to the nucleon density operators characterized by coupling constants $f_p$ and $f_n$ to protons and neutrons, respectively. In a wide range of BSM scenarios~\cite{Jungman:1995df, Arrenberg:2008wy}, the Higgs-to-strange quark coupling is the dominant component of the $\chi$$A$ interaction. Since the proton and neutron have similar strange quark contents, it is usually assumed that $f_p \simeq f_n$. The $\sigma_{\chi A}$ is, therefore, simplified to be proportional to $A^2$. This {\it $A^2$}--scaling of the cross section is a very strong driving force in the direct detection of dark matter experiments and is analogous to the {\it $(A-Z)^2$}--scaling in CENNS. Moreover, it has been known that the CENNS of astrophysical and atmospheric neutrinos are irreducible backgrounds for future generation dark matter detectors at spin-independent cross-sections. A recent study showed background limits to future dark matter searches coming from CENNS interactions of astrophysical and atmospheric neutrinos~\cite{Billard:2013qya}. There are a few possible ways to improve the limits by using directional measurements of the neutrino interactions and/or measuring time variation of the interactions. However, this CENNS background limit is a robust lower bound which can not be substantially reduced.  Measuring the CENNS cross section and performing subsequent tests of higher energy neutrino interactions on various target materials will be extremely beneficial to future dark matter experiments. The importance of the CENNS physics cases in dark matter searches is also pointed out in a recent Snowmass report~\cite{Cushman:2013zza}.\par

%-------------------------------------------------------------------------------------
\subsubsection{Supernova Physics}
%-------------------------------------------------------------------------------------
	The major unsolved problem of a supernova explosion is to understand how the burst of neutrinos transfers its energy to produce the shock wave that causes the star to explode. CENNS plays a major role in an explosion of a core-collapse supernova~\cite{Freedman:1977xn}. In the core of the dying star, neutrinos are scattered, absorbed, and reemitted by super-dense proton-neutron matter. Although yet to be fully understood, modern numerical simulations show that neutrino-driven convection eventually causes the giant star to explode. A CENNS cross-section different from the nominal SM prediction could have significant impact on the understanding of supernova explosions. \par	
	Moreover, CENNS is an important process for the detection of supernova neutrinos. Future large-scale low-energy threshold underground detectors, such as the CLEAN detector~\cite{Nikkel:2006nh}, will be sensitive to all active neutrino species in a supernova burst, and will be flavor blind~\cite{Horowitz:2003cz}. Hence, detecting supernova neutrinos in such a detector may provide a total flux and spectrum of neutrinos from supernova if the cross section of CENNS can be independently and accurately measured. These results combined with flavor-dependent interaction measurements~\cite{Scholberg:2012id,Chakraborty:2013zua} can explain how neutrinos are thermalized with matter in a supernova. \par

%=====================================================================================
\subsection{CENNS in Particle Physics}
%=====================================================================================
%-------------------------------------------------------------------------------------
\subsubsection{Neutrino Oscillations}
%-------------------------------------------------------------------------------------
	Neutrino flavor oscillation is a well established physics phenomenon studied over the last four decades. Neutrino disappearance and appearance signatures are successfully explained by representing the neutrino flavor eigenstates as a mixture of non-zero mass eigenstates. There has been huge progress in measuring neutrino mixing angles during the last decades. Identifying mass hierarchies, measuring CP-phase(s) and determining whether neutrinos are Dirac or Majorana particles are active topics in the field. CENNS is a large and well-predicted cross-section in the Standard Model. If discovered at its predicted rate, the CENNS process can become a powerful tool for future low energy neutrino physics, especially for neutrino oscillation experiments.\par

	A number of recent anomalous results suggest the existence of a sterile neutrino~\cite{Aguilar:2001ty, Aguilar-Arevalo:2013pmq}. In these experiments, an excessive appearance of active-flavor neutrinos is seen. If confirmed, this excess requires a model which has relatively large mass differences ($\Delta m^2 \sim 1$~eV$^2$) and requires at least one more mass eigenstate ($m_4$) in the neutrino mass spectrum. Most of the previous experiments are based on charged-current measurements, and hence are indirectly inferring the mixing matrix elements.  However, the sterile neutrino models can be clearly verified by CENNS interactions. The CENNS interaction is insensitive to the differences of active flavors of neutrinos, thus the measurement will be of total fluxes of active flavor neutrinos. Sterile neutrino oscillations manifest themselves as a baseline- and energy-dependent {\it disappearance} of CENNS interactions. A short-baseline neutrino experiment measuring CENNS has the potential to probe a wide range of oscillation hypotheses~\cite{Garvey:2005pn,Anderson:2012pn}. \par
	
	A sensitivity study of a future sterile neutrino search using CENNS has been carried out in reference~\cite{Anderson:2012pn}. The study assumes neutrino fluxes of $2.5\times10^7 (6.3\times10^6) \nu$/cm$^2$/sec per flavor at 20\,m (40\,m) from the pion decay-at-rest neutrino source with one near (20\,m) detector with 456\,kg of liquid argon and four far (40\,m) detectors. With this experimental scenario, one can test the LSND best-fit mass splitting ($\Delta m^2=1.2\,$eV$^2$) at the 3.4 sigma.

%-------------------------------------------------------------------------------------
\subsubsection{Neutrino Magnetic Moment}
%-------------------------------------------------------------------------------------
%FIG==============
\begin{figure}[t]
\centering
\includegraphics[width=3.5in]{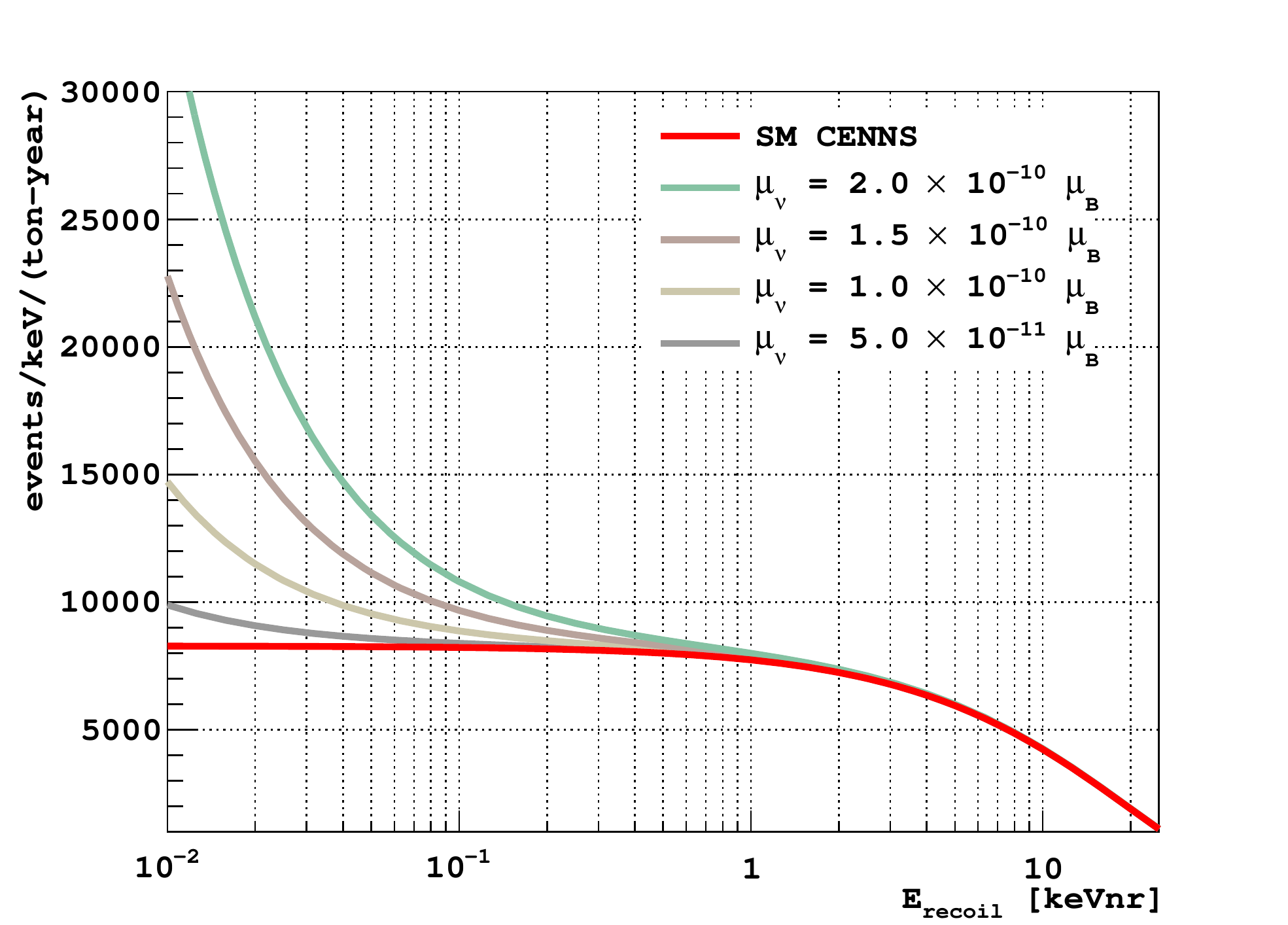}
{\caption{{Differential yield as a function of nuclear recoil energy for different values of neutrino magnetic moment ($\mu_\nu (\nu_\mu)$). $\nu_{\mu}$ flux of $2.5\times 10^7~\nu$/cm$^2$/s from pion decay-at-rest source is assumed.}}\label{fig:magneticmom}}
\end{figure}
%FIG==============

As a consequence of non-zero masses, neutrinos can have magnetic moments. In the minimally extended SM, Dirac neutrinos of mass $m_\nu$ have a magnetic moment through one-loop radiative corrections~\cite{Vogel:1989iv}. The magnetic moment is given by

\begin{equation}
\mu_\nu =\frac{3 G_F m_e m_\nu}{4\pi^2 \sqrt{2}}\mu_B \simeq 3.2 \times 10^{-19}\left(\frac{m_\nu}{1~\mbox{eV}}\right)\mu_B,
\end{equation}

\noindent where $G_F$ is Fermi constant, $m_e$ is electron mass, and $\mu_B(=e/2m_e)$ is Bohr magnetons. This predicted value in an extended SM is exceedingly small to be measured. However, beyond the SM (BSM) models commonly predict larger values of $\mu_\nu$, and hence any measurement of excessive neutrino magnetic moment would be a signature of BSM physics~\cite{Bell:2006wi}. There are several consequences of the neutrinos having large magnetic moments. The neutrino-electron scattering cross section would be modified in low energies. Neutrinos would flip their spin in  strong external magnetic fields which is, for example, a natural configuration for the core region of stars. Heavier-mass neutrinos would decay radiatively to lighter-mass neutrinos and emit photons.\par

The best direct experimental limits result for $\nu-e$ scattering is from GEMMA experiment, $\mu_\nu(\bar{\nu}_e) \le 0.32 \times 10^{-10}\mu_B$~\cite{Beda:2009kx}. For muon neutrino scattering, the best limit is less stringent: $\mu_\nu(\nu_\mu) \le 6.8 \times 10^{-10}\mu_B$~\cite{PhysRevD.63.112001}. The most stringent limits are from astrophysical observations with several assumptions. For example a model-dependent analysis of plasmon decay in red giant evolution~\cite{PhysRevLett.64.2856}, and an analysis of neutrino spin-flip precession in Supernova 1987A set limits of $\mu_\nu \le 10^{-12}\mu_B$~\cite{PhysRevD.59.111901}.\par

A finite neutrino magnetic moment can be observed in the recoil spectrum of CENNS. The magnetic scattering cross section is given by~\cite{Vogel:1989iv},

\begin{equation}
\left(\frac{d\sigma}{dE_R}\right)_m = \frac{\pi \alpha^2 \mu_\nu^2 Z^2}{m_e^2}\left(\frac{1-E_R/E_\nu}{E_\nu}+ \frac{E_R}{4E_\nu^2} \right),
\end{equation}

\noindent where $\alpha$ is the fine structure constant, $E_R$ is the recoil energy of nuclei. FIG.~\ref{fig:magneticmom} shows the event rates as a function of energy thresholds in a germanium detector with pion decay-at-rest $\nu_{\mu}$ flux of $2.5\times 10^7 \nu$/cm$^2$/s for various magnetic moment contributions. Future detectors for sub-keV recoil energy thresholds would begin to directly test new regimes of neutrino magnetic moment.

%-------------------------------------------------------------------------------------
\subsubsection{Non-Standard Model Interactions}
%-------------------------------------------------------------------------------------
%FIG==============
\begin{figure}[ht!]
\centering
\includegraphics[width=3.5in]{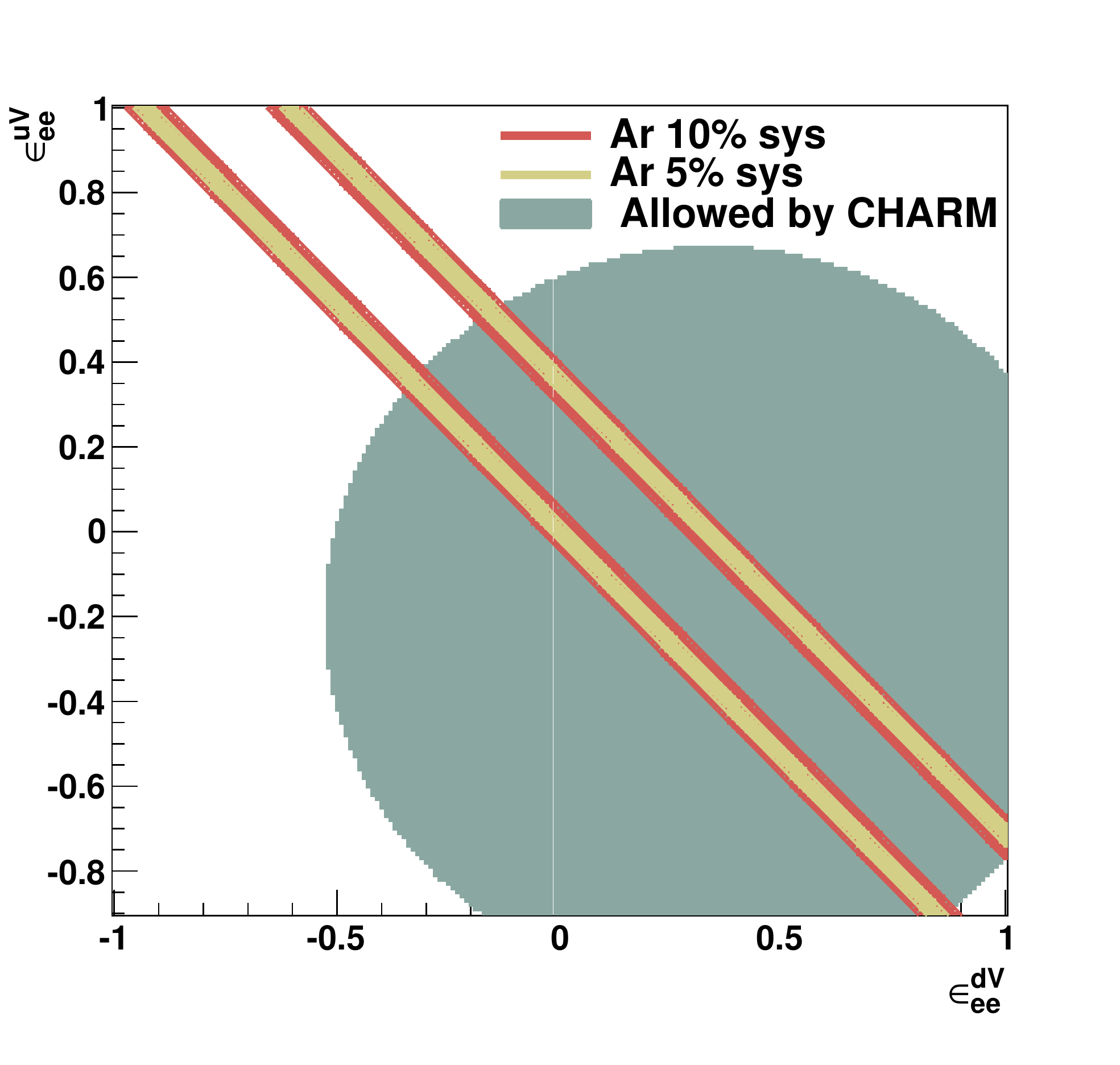}
{\caption{ {Allowed regions (red and yellow shaded areas) at 90\% C.L. assuming measurement of the SM-predicted CENNS rate, for $\epsilon^{\mu V}_{ee}$ and $\epsilon^{dV}_{ee}$ for 1\,ton-year liquid argon detector at $5\times10^6 \nu/$cm$^2$/s per flavor of pion decay-at-rest neutrino flux, assuming 5\% or 10\% of systematic uncertainty in measurement. The energy threshold is assumed at 25\,keV$_{nr}$ (nuclear recoil). The shaded elliptical region corresponds to a slice of the CHARM-experiment-allowed NSI parameter space, for $\epsilon^{qA}_{ee}$=0.}}\label{fig:nsi}}
\end{figure}
%FIG==============

CENNS is a well-predicted cross-section in the Standard Model. Therefore any deviation from the predicted value would be an indication of BSM physics. Any non-standard interactions (NSI) which are specific to the interactions of neutrinos and quarks can be parameterized in a relatively model-independent way. An effective Lagrangian of a neutrino with a hadron in the parametrization of $\epsilon_{ij}$ can be described as ~\cite{Barranco:2005yy,Davidson:2003ha};
\begin{eqnarray}
\mathcal{L}^{NSI}_{\nu H} & = & - \frac{G_F}{\sqrt{2}}{\displaystyle \sum_{{q=u,d}\atop{\alpha, \beta = e, \mu, \tau}}} [ \bar{\nu}_\alpha \gamma^\mu (1-\gamma^5) \nu_\beta] \\ \nonumber
	&&\times (\varepsilon_{\alpha \beta}^{qL}[ \bar{q} \gamma_\mu (1-\gamma^5) q]+ \varepsilon_{\alpha \beta}^{qR} [ \bar{q} \gamma_\mu (1+\gamma^5) q]),
\end{eqnarray}

\noindent where the $\varepsilon$ parameters represent either {\it non-universal} ($\alpha = \beta$) or {\it flavor-changing} ($\alpha \ne \beta$) interactions. Many of these parameters are quite poorly constrained, and CENNS experiments can improve sensitivity by an order of magnitude ~\cite{{Barranco:2005yy},{Scholberg:2005qs},{Barranco:2007tz}}. The cross section for CENNS of $\nu_\alpha$ off a spin-zero nucleus (A) is given by

\begin{eqnarray}
\begin{aligned}
& \left(\frac{d\sigma}{dE}\right)_{\nu_\alpha A} = {\displaystyle \frac{G_F^2 M}{\pi} F^2(2ME)}\left[1 - \frac{M E}{2k^2}\right]  \times \nonumber \\
& {\displaystyle  \{[Z(g_V^p + 2\varepsilon_{\alpha \alpha}^{uV}+ \varepsilon_{\alpha \alpha}^{dV})+ N(g_V^{n} + \varepsilon_{\alpha \alpha}^{uV}+ 2\varepsilon_{\alpha \alpha}^{dV})]^2} \nonumber \\ 
& + \displaystyle \sum_{\alpha \ne \beta }{[Z(2\varepsilon_{\alpha \beta}^{uV}+ \varepsilon_{\alpha \beta}^{dV})+ N(\varepsilon_{\alpha \beta}^{uV} + 2 \varepsilon_{\alpha \beta}^{dV})]^2 \}},
\end{aligned}
\end{eqnarray}\label{eq:nsi}

\noindent where $g_V^p = (\frac{1}{2}-2\sin^2 \theta_W)$, $g_V^n=-\frac{1}{2}$ are the SM weak constants. FIG.~\ref{fig:nsi} shows allowed regions for $\epsilon^{\mu V}_{ee}$ and $\epsilon^{dV}_{ee}$, for 1\,ton-year of liquid argon detector data assuming high-intensity pion decay-at-rest neutrino flux. The shaded elliptical region corresponds to constraints by the CHARM experiment~\cite{Dorenbosch:1986tb}. Hence a CENNS experiment at an intense stopped-pion neutrino source would have significant sensitivity to currently-allowed NSI interaction parameters.

%=====================================================================================
\subsection{CENNS in Nuclear Physics}
%=====================================================================================
Determination of the neutron distributions in nuclei is important not only for fundamental understanding of nuclear physics, but also because of important implications for astrophysics. For example, the primary physics quantities of neutron stars such as masses, radii, and composition are determined using the equations of states of neutron-rich nuclei. The equation of state is related to the nuclear symmetry energy, which is defined as, $E(n,\delta) \simeq E_0(n)+E_{\mbox{sym}}\delta^2$ and $\delta = (n_n - n_p)/(n_n+n_p)$, where $n_n$ and $n_p$ are the number densities of neutrons and protons. The symmetry energy is strongly correlated with the skin thickness of neutrons~\cite{Steiner2005325}, and hence the radii of neutrons. Therefore, the size of neutron stars can be predicted more precisely based on better measurements of the equation of state. Traditional methods of measuring neutron radii through hadronic scattering report typical uncertainties of order 10\%~\cite{Patton:2012jr}. \par

The CENNS interaction is especially sensitive to neutron numbers in target nuclei, which provides a clean way to measure the neutron part of nuclear form factors. At low momentum transfer, the form factor $F(Q^2)\sim 1$. However for higher $Q$ values, small deviations from coherence occur as higher-order terms of the nuclear form factors come into play~\cite{Amanik:2009zz}. The CENNS cross section of the spin-zero nuclears is given by,

\begin{equation}
\left(\frac{d\sigma}{dE}\right)_{\nu A} = \frac{G_F^2}{2\pi}\frac{Q_w^2}{4} F^2(Q^2) M \left[2 - \frac{M E_R}{E_{\nu}^2}\right], 
\end{equation}

\noindent where $M$ is the nuclear mass, $Q_w$ is the weak charge and $Q=\sqrt{2ME_R}$. The form factor $F(Q^2)$ can be expanded as:

\begin{equation}
F_{n}(Q^{2}) \sim N \left( 1 - \frac{Q^{2}}{3!} \langle R^{2}_{n} \rangle + \frac{Q^{4}}{5!}\langle R^{4}_{n}\rangle + \cdots \right)\,,
\end{equation}\label{eqn:formfactor1}

\noindent where $\langle R^i_n \rangle$ are the even moments of the neutron density. Such deviations are observable as small distortions of the expected recoil spectral shape and can be exploited to measure nucleon density distributions.  With good control of spectral shape uncertainties, multi-ton-scale experiments could make meaningful measurements of the neutron radius $\langle R_n^2 \rangle^{1/2}$ and potentially higher-order moments.

According to reference~\cite{Patton:2012jr}, a exposure of 3.5\,ton-year with a liquid argon detector with neutrino flux of $3\times10^7 \nu/$cm$^2$/s per flavor is required to measure the 2nd and 4th moments of the form factor. The experimental requirements are challenging to reach in the near future; however, it is possible to determine the neutron radius to a few percent by measuring neutron form factor with sufficient accuracy. The precise measurements of neutron radii then improve the predictive power of the equation of state of neutron matter, and thus the knowledge of the size of neutron stars~\cite{Steiner2005325,Horowitz:2000xj}.

%=====================================================================================
\subsection{Summary}
%=====================================================================================
In order to achieve the above physics goals, a phased approach is most appropriate, depending on the available neutrino beam power and detector technology. 

\begin{enumerate}

\item The first-generation CENNS experiment would be the discovery of the CENNS interaction and measure the cross-section with $\sim$10\% of accuracy, for example at Fermilab. The experiment can be carried out with existing dark matter detector technology, existing beamline and target station at Fermilab. The result would be sensitive to the NSI ranges as well. 

\item The second-generation experiment would be the precision measurement of the CENNS cross-section. The accurate measurement of the neutrino flux, assuming cross section is exactly known, would be a powerful tool for neutrino oscillation study~\cite{Elnimr:2013wfa} and future low energy neutrino experiments. This would also allow an initial series of measurements of supernova-related neutrino cross sections on a variety of targets~\cite{Efremenko:2008an}, where they have rarely been measured. The precision measurement of the CENNS cross section will be a valuable input to the next generation of dark matter experiments. 

\item The third-generation CENNS experiment would use a high-intensity neutrino beam and large-scale neutrino detector with a lower energy threshold. The goal would be a search for the neutrino magnetic moment, measurement of the neutron form factor, and search for possible deviations of the SM. 

\end{enumerate}

The major focus in this paper is the first-generation of CENNS experiment  -- the discovery of the CENNS. There are a few existing intensive pion decay-at-rest neutrino sources, for example the Spallation Neutron Source at Oak Ridge National Laboratory~\cite{Bolozdynya:2012xv}. In this paper we present an alternative promising setup at an existing neutrino beam at Fermilab. This is a unique idea of using the neutrinos at a location off-axis of existing high energy neutrino beam~\cite{Yoo:2011}. \par

%%%%%%%%%%%%%%%%%%%%%%%%%%%%%%%%%%%%%%%%%%%%%%%%%%%%%%%%%%%%%%%%%%%%%%%%%%%%%%%%%%%%%%%
\section{Low Energy Neutrino Source at Fermilab}\label{bnb}
%%%%%%%%%%%%%%%%%%%%%%%%%%%%%%%%%%%%%%%%%%%%%%%%%%%%%%%%%%%%%%%%%%%%%%%%%%%%%%%%%%%%%%%
%FIG==============
\begin{figure*}[th!]
\centering
\includegraphics[width=6.5in]{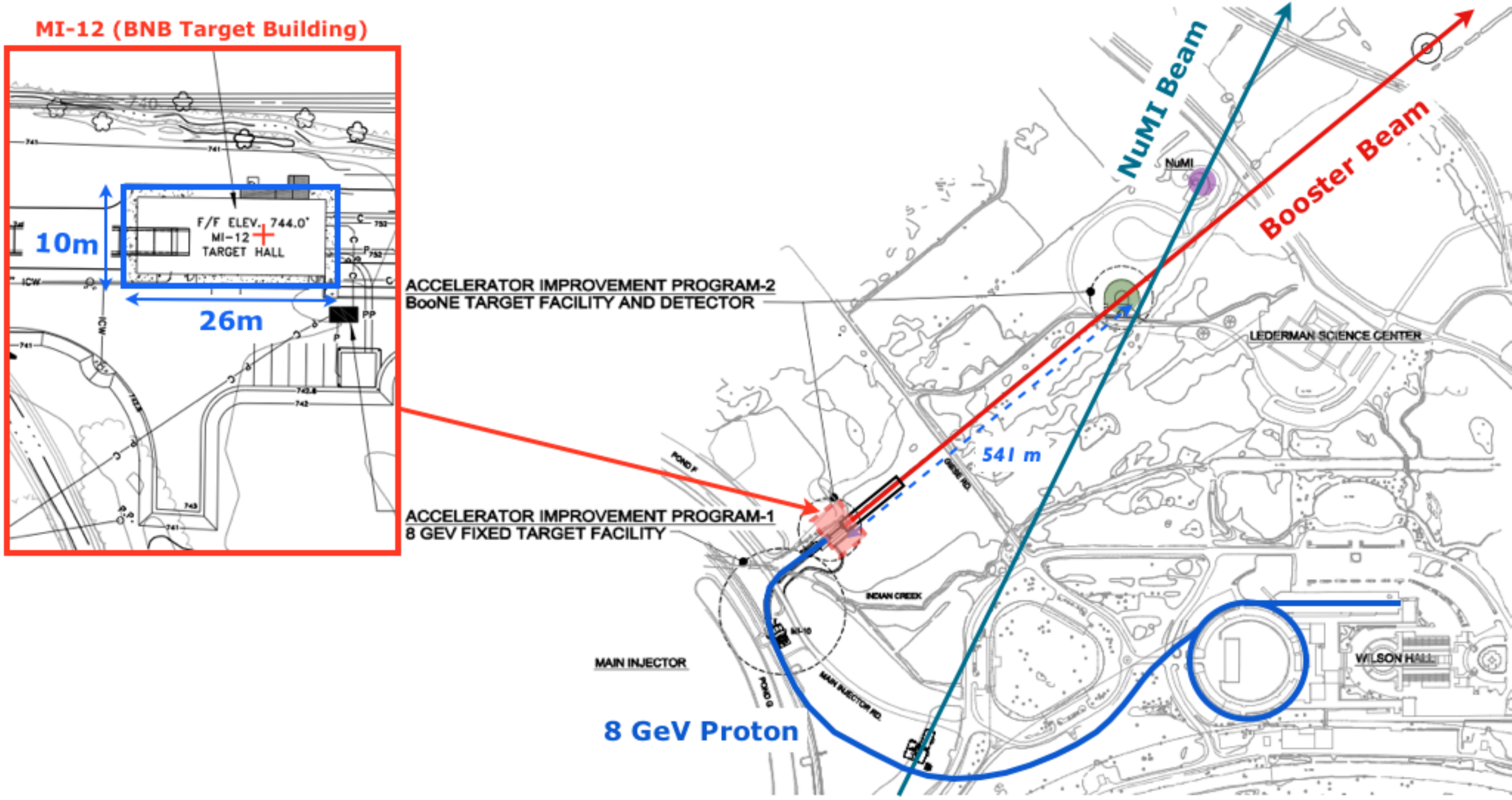}
{\caption{ Fermilab neutrino beam lines: the Booster Neutrino Beam (BNB, red line) and Neutrinos at the Main Injector beam (NuMI, green line)~\cite{bnbmap:1999}. The left insert figure shows the configuration around BNB target building (MI-12) area~\cite{BNBDRAWING:2000}. The red cross in the figure indicates the location of the target. No facility equipment occupies the area near the potential experiment site.}\label{fig:bnbmap}}
\end{figure*}
%FIG==============

	Fermilab has two major neutrino beamlines (see FIG.~\ref{fig:bnbmap}); the Neutrinos at the Main Injector (NuMI) and the Booster Neutrino Beam (BNB). The energy range of these two neutrino sources on-axis is in the GeV range, which is too high to satisfy the condition for dominance of coherent scattering. We found the far-off-axis ($>$ 45 degrees) of the BNB produces well defined neutrinos with energies below $\sim$50\,MeV. The BNB source has substantial advantages over the NuMI beam source owing to suppressed kaon production from the relatively low energy 8\,GeV proton beam on the target. Therefore, pion decay and subsequent muon decay processes are the dominant sources of neutrinos. At the far-off-axis area, the detector can be placed close enough to the target to gain a large increase in neutrino flux due to the larger solid angle acceptance. An initial study using the existing BNB MC has confirmed that this approach is promising.\par

	The Fermilab Booster is a 474-meter-circumference synchrotron operating at 15\,Hz. Protons from the Fermilab LINAC are injected at 400\,MeV and accelerated to 8\,GeV kinetic energy. The structure of the beam is a series of 81 proton bunches each with a 2 ns width and 19\,ns apart. The maximum average repetition rate for proton delivery to the BNB is 5 Hz and $5\times10^{12}$ protons per pulse. The repetition limit is set by the horn design and its power supply. The target is made of beryllium divided in seven cylindrical sections in a total of 71.1\,cm in length and 0.51\,cm in radius. In order to minimize upstream proton interactions, the vacuum of the beam pipe extends to about 152\,cm upstream of the target. The horn is an aluminum alloy toroidal electromagnet with operating values of 174 kA and maximum field value of 1.5 Tesla. A concrete collimator is located downstream of the target and guides the beam into the decay region. The air-filled cylindrical decay region extends for 45 meters. The beam stop is made of steel and concrete. Details of the Fermilab BNB neutrino fluxes can be found in~\cite{AguilarArevalo:2008yp}.\par

	At very far-off-axis the pion decay region is no longer a point source and the angle from on-axis is not a well defined quantity. Moreover, the geometry around the target area and shielding should be properly taken into account in the neutrino flux calculation as the secondary hadronic processes in the shielding material also produce pions and hence neutrinos. In particular, the pion production from the 8\,GeV Booster proton beam on the beryllium target, the multiplicity of pion production from the subsequent $\pi$-$p$ interactions and defocused $\pi^{-}$s from the horn all require a well modeled MC study. \par

	In order to understand the neutrino flux at BNB far-off-axis, we adapted the Booster Neutrino Beam Monte Carlo (BNB MC). The BNB MC uses the Geant-4 framework for propagating particles, for electromagnetic processes, hadronic interactions in the beamline materials and the decay of particles. The geometry of the target area and beamline is accurately modeled. The double differential cross sections of pion and kaon production in the simulation have been tuned to match external measurements. This is true for the hadronic cross sections for nucleons and pions as well~\cite{AguilarArevalo:2008yp}. The original BNB MC, however, contains a hard-coded tracking threshold cut to remove stopping pions (defined as below 1\,MeV in kinetic energy). In fact, the stopping pions are the dominant neutrino source at far-off-axis. The cut does not affect any previous on-axis Booster Beam experiments such as MiniBooNE and SciBooNE which focus on above-100-MeV neutrino interactions.\par

%FIG==============
\begin{figure}[t]
\centering
\includegraphics[width=3.15in]{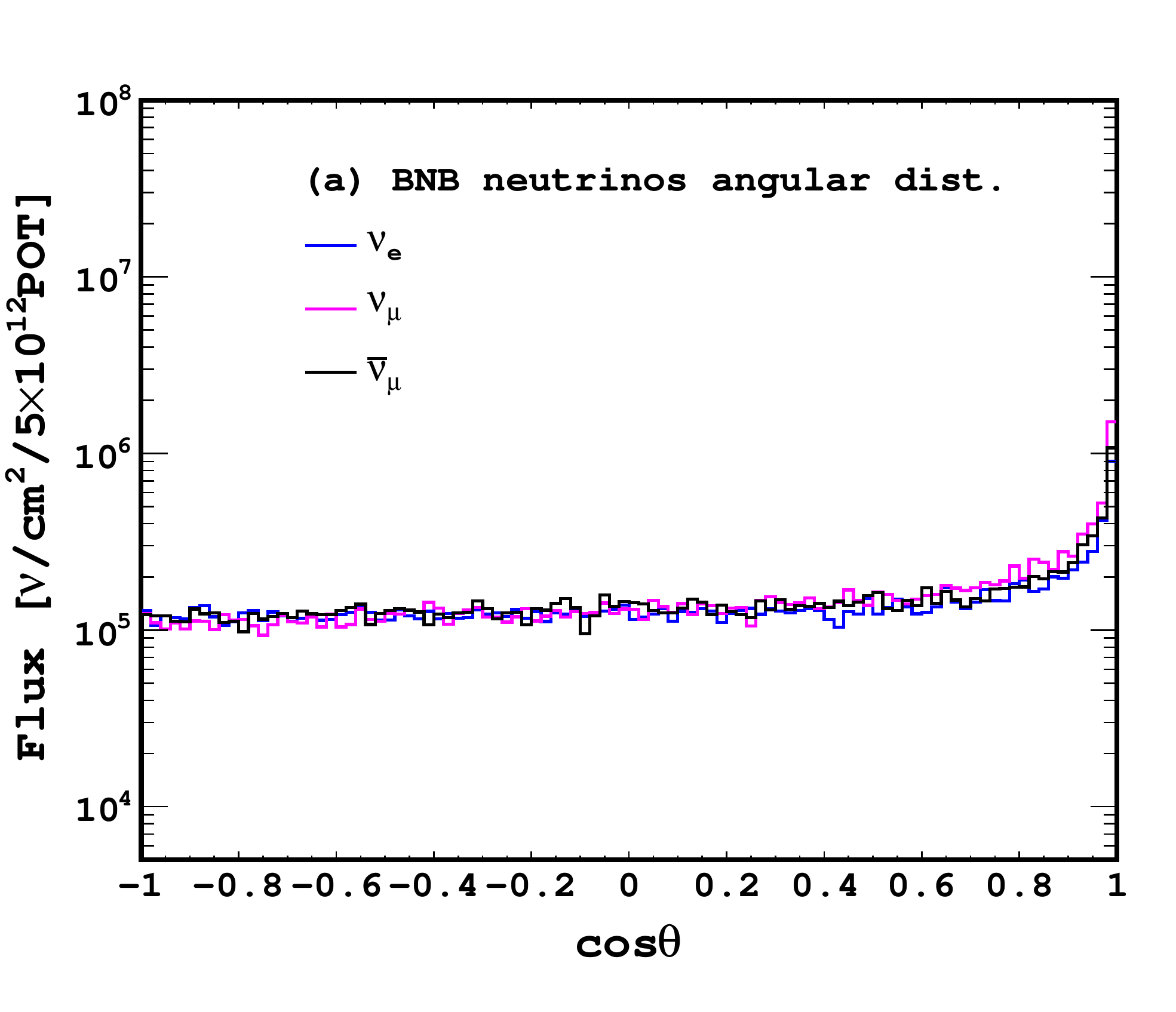}
\hspace{0.2cm}
\includegraphics[width=3.15in]{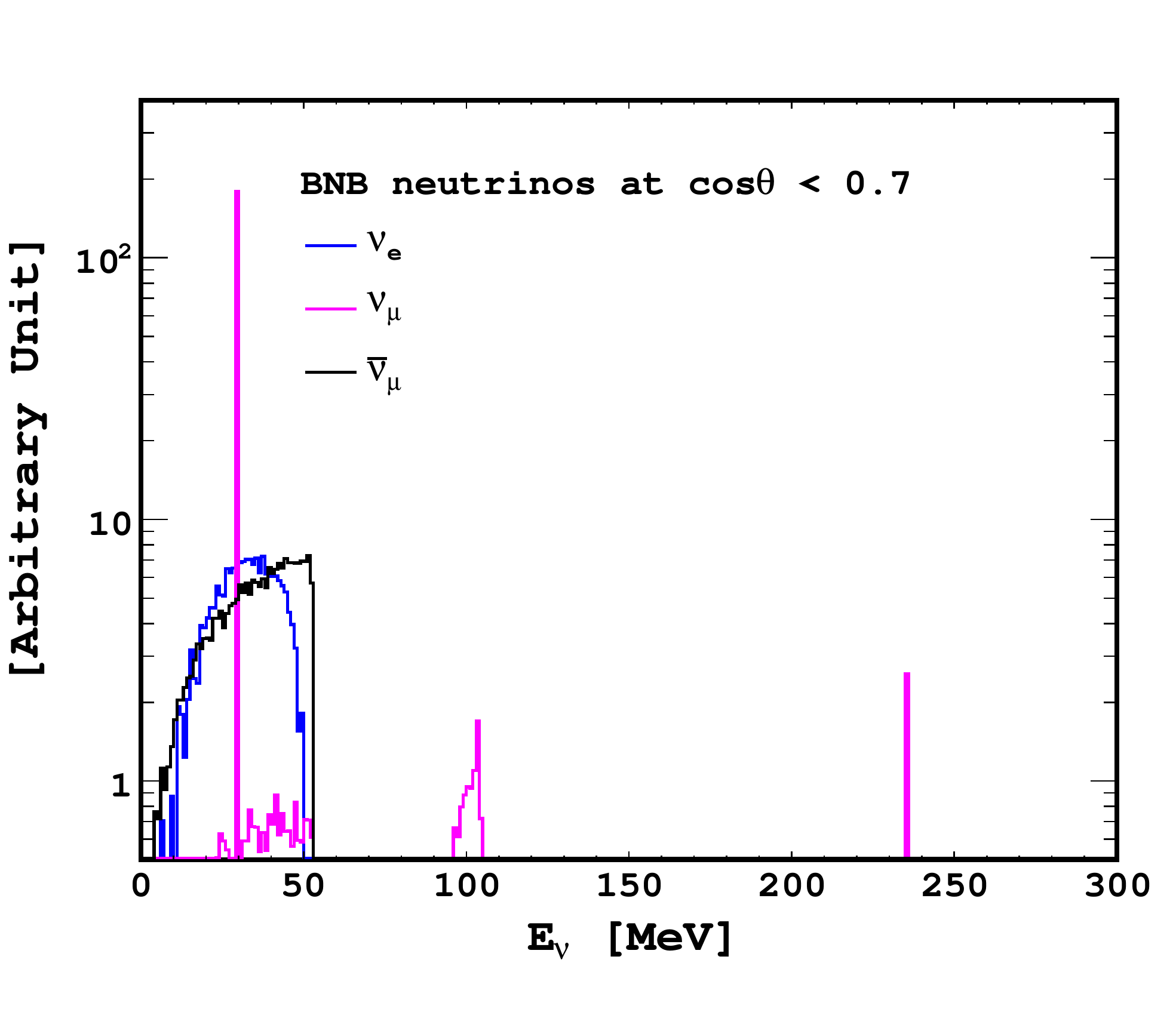}
{\caption{ Estimated neutrino flux from modified BNB MC in $\nu$-mode, 173kA horn current and 8 GeV booster beam configuration. The neutrino flux is normalized per $5\times10^{12}$ protons on target. (a) The angular dependence of the neutrino fluxes for different flavors. The flux becomes uniform below $\cos\theta<0.7$. See text for the definition of $\theta$. (b) Energy spectrum of neutrinos below $\cos\theta<0.7$ (far-off-axis) for different flavors.}\label{fig:bnbangspc}}
\end{figure}
%FIG==============
	
	The BNB MC simulation was carried out in neutrino mode with 173\,kA horn current and 8\,GeV proton momentum. FIG.~\ref{fig:bnbangspc}(a) shows the angular distribution of the neutrino flux 20\,m away from a reference point of the upstream end of the decay pipe where the angle is measured from on-axis. The flux of the neutrinos, at the 32\,kW maximum Booster power ($5\times10^{12}$ protons on target (POT) per pulse), is estimated to be about $10^5 \nu$/cm$^2$/pulse per flavor with 5\,Hz frequency within a pulse width of 1.6\,$\mu$s. Hence, the neutrino flux per unit time is about 5$\times 10^5 \nu$/cm$^2$/s.  FIG.~\ref{fig:bnbangspc}(b) shows the energy spectrum of neutrinos at angles less than $\cos\theta <0.7$ which is dominated by neutrinos from stopping pion decay. The pion decay at rest ($\pi^+ \rightarrow \mu^+ \nu_\mu$) produces a prompt and monochromatic $\nu_\mu$ at 29.9\,MeV. The $\mu^+$ then decays on a 2.2\,$\mu$s timescale to produce a $\bar{\nu}_\mu$ and a $\nu_e$ with energies between 0 and $m_\mu/2$. In FIG.~\ref{fig:bnbangspc}(b), the $\nu_\mu$, $\nu_e$ and $\bar{\nu}_\mu$ spectra follow the stopping $\pi^+$ decay kinematics. The small $\nu_\mu$ bump at $\sim$100\,MeV is due to the neutrinos from $\mu^-$ capture on nuclei. The peak at 235.3\,MeV is from kaon decay at rest. These $\nu_\mu$s above 55\,MeV are potential background sources since the interaction of neutrinos may scatter off neutrons from nuclei nearby or inside the detector.\par 
	
\begin{figure}[t!]
\centering
\includegraphics[width=3.5in]{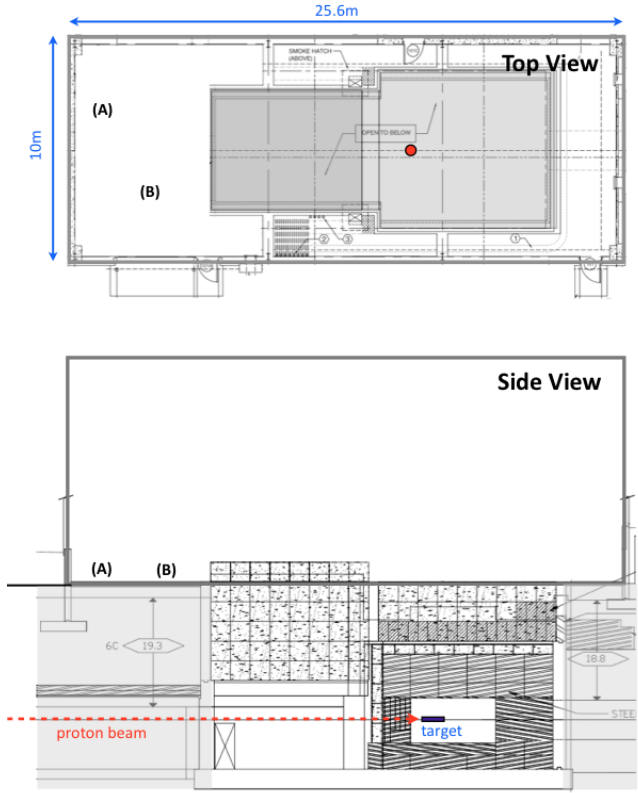}
{\caption{ The top-down and elevation views of the BNB target building. The SciBath detector was operated at location (A) and the EJ-301 measurement carried out at location (B). The drawing is taken from~\cite{BNBDRAWING:2000} and modified. The red-filled circle in the top figure indicates the upstream end of the target position. }\label{fig:mi12shielding}}
\end{figure}

	The existing radioactive shielding at the BNB target area is extensive and carefully thought out in order to satisfy the Fermilab radioactive safety regulations~\cite{BNBTDR:2001} (see FIG.~\ref{fig:mi12shielding}). The target itself is located $\sim$7\,m underground from the building surface. The shielding pile consists of iron blocks totaling 2.6\,m in elevation (1,600\,tons), an additional 3.2 m-thick concrete shielding (300\,tons), and special custom sized steel (40\,tons) above and below the horn module. About 3$\times$10$^{22}$ neutrons per 10$^{21}$\,POT (year) are expected to be initially produced at the target. These neutrons are produced in the forward beam direction with a maximum kinetic energy of $\sim$8\,GeV with more than 90\% of neutrons below 50\,MeV. The high energy neutrons scatter off the surrounding materials and produce secondaries. Considering the existing shielding configuration, the beam-induced neutron flux at about 20\,m away from the target is roughly estimated to be $\sim$3.6$\times$10$^{8}$\,neutrons/$m^2$ per 10$^{21}$POT. According to a simple scaling of neutron shielding, an additional $\sim$8\,m-thick concrete barrier would be enough to shield out most of the beam-induced neutrons. Although the estimated beam-induced neutron background is sufficiently low, it is also true that predicting neutron leakage rates through massive shielding material is notoriously difficult. For example, a small gap between shielding blocks may potentially cause serious leakage of neutron fluxes. Fast neutron background, if there is any, would require a more extensive study for the shielding design. Therefore, measuring the beam coincident neutron flux and energy spectrum at the experimental site is necessary. With the help of the Fermilab Accelerator Division, we carried out beam-induced background studies at the BNB target building, which is described in section~\ref{neutronbg}.

%FIG==============
%FIG==============
	
	The far-off-axis site of the BNB is also the far-off-axis site of the NuMI beam (see FIG.~\ref{fig:bnbmap}). The NuMI beam contains a potential source of background from high energy neutrinos ($>$200\,MeV) from kaon decay. However, the distance from the NuMI target to the BNB far-off-axis site is more than 200\,m away and the NuMI neutrinos can be vetoed out using beam trigger information. Therefore the neutrinos from the NuMI beamline should be significantly suppressed. \par 

	Beam-uncorrelated backgrounds are mitigated by the BNB beam window; the timing allows a factor of 5$\times 10^{-5}$ rejection ({\it duty factor}, here after) of steady-state backgrounds assuming a 10\,$\mu$s detector time window. The total detector beam-on livetime per year is only $\sim$26\,min(=5$\times 10^{-5}\times$year). Timing of individual events in the detector can be known to within $\sim$10\,ns using detectors with fast timing. Furthermore, these backgrounds can be subtracted using beam-off data. Cosmic ray-related backgrounds will be significantly reduced by the water shielding veto system.\par

%%%%%%%%%%%%%%%%%%%%%%%%%%%%%%%%%%%%%%%%%%%%%%%%%%%%%%%%%%%%%%%%%%%%%%%%%%%%%%%%%%%%%%%
\section{Neutron Backgrounds Measurement}\label{neutronbg}
%%%%%%%%%%%%%%%%%%%%%%%%%%%%%%%%%%%%%%%%%%%%%%%%%%%%%%%%%%%%%%%%%%%%%%%%%%%%%%%%%%%%%%%

A commercial EJ-301 liquid scintillator neutron detector and a newly-developed neutral particle detector, named {\it SciBath}~\cite{Tayloe:2006ct, Cooper:2011kx} were used to measure the neutron backgrounds in the BNB target building. 

%=====================================================================================
\subsection{EJ-301 liquid scintillator}
%=====================================================================================
%FIG==============
\begin{figure}[t!]
  \centering
  \includegraphics[width=3.6in]{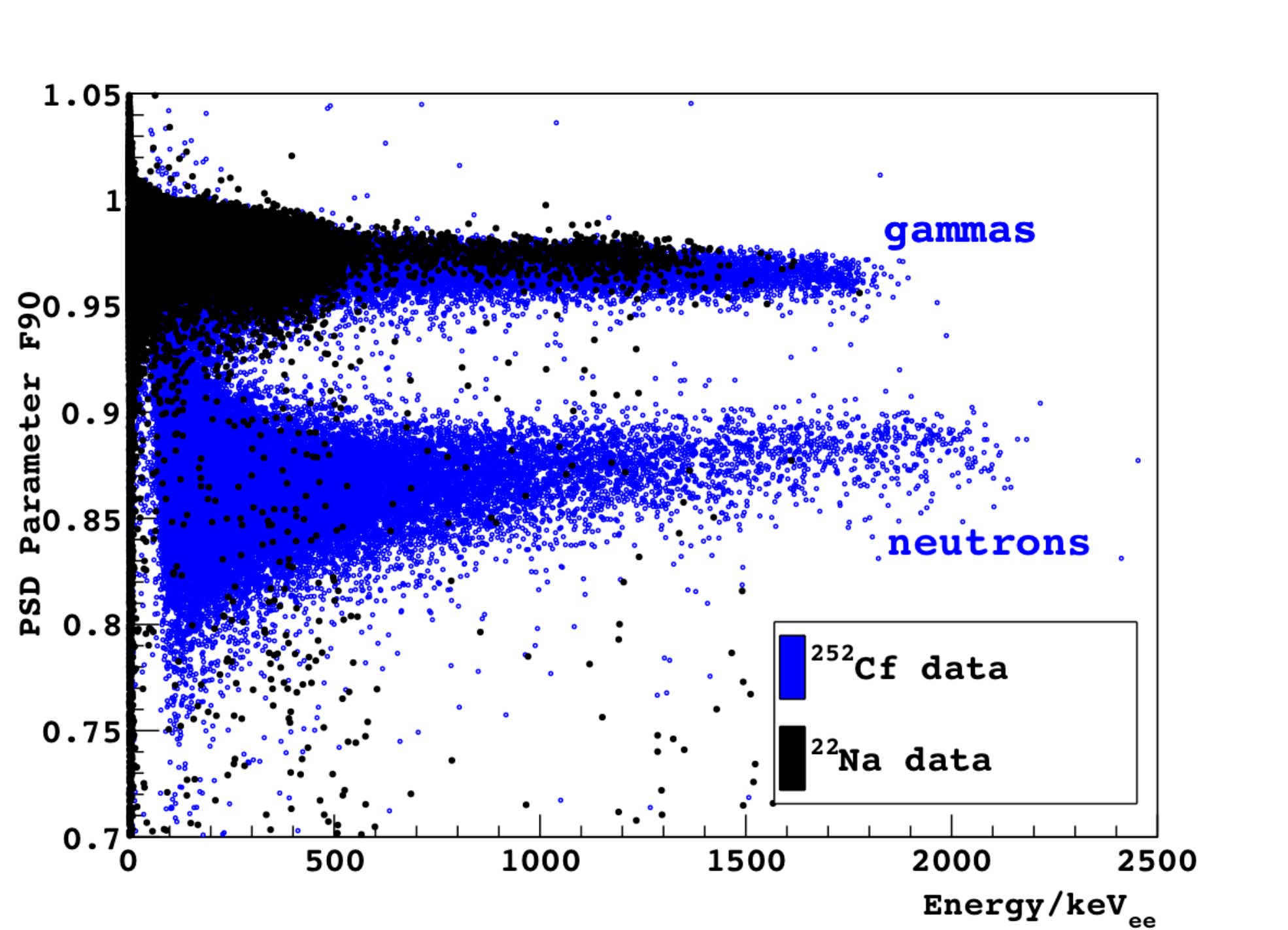}
  \caption{Calibration of the energy and pulse shape discrimination (PSD) parameter F90 for the EJ301 scintillator detector with neutron ($^{252}$Cf) and gamma ($^{22}$Na) sources. The 511 keV annihilation gamma is easily visible in the $^{22}$Na data. Neutrons are defined as events having F90 between 0.76 and 0.91 and energy below the digitizer saturation point (around 2\,MeV$_{ee}$, electron equivalent energy, on this scale; saturating events are excluded from this plot) and above the point where the gamma and neutron F90 distributions merge (around 200\,keV$_{ee}$ on this scale).}
  \label{fig:ej301_psdcalibration}
\end{figure}
%FIG==============

To obtain a rough estimate of the neutron background from the Booster beam, we attempted to measure the neutron flux with a commercial liquid scintillator detector (Eljen 510-50x50-1/301 Liquid Scintillation Detector Assembly, sealed system with 5'' ETEL-9390KB PMT and EJ301 scintillator). The PMT signals were recorded from 3\,$\mu$s before to 20\,$\mu$s following the beam trigger using a CAEN V1720 250\,MS/s, 12-bit, 2\,Vpp digitizer. The scintillation response of the cell to gammas of various energies was calibrated using the Compton edges of $^{133}$Ba, $^{137}$Cs, and $^{22}$Na sources, from which the energy of proton recoils can be obtained using Table~1 of reference~\cite{Verbinski}. (The scintillation light output for 1\,MeV proton recoils is quenched by a factor $\sim$0.16 relative to electrons.)  Given the gain of the phototube, pulses begin to exceed the vertical range of the digitizer at around 500 p.e. (photoelectron) for gamma events ($\sim$700\,p.e. for neutron events due to the slower scintillation pulses). We have measured beam-induced events with energies up to 8000 p.e., or $>$4\,MeVee, but we have not calibrated the effect of the digitizer saturation in order to correct the energy scale at these energies.

Discrimination between electron recoil (gamma-induced) and nuclear recoil (neutron-induced) events can be achieved via pulse shape discrimination (PSD)~\cite{Adams,Cecil}. We have adopted F90, the fraction of photons collected in the first 90 ns of a scintillation pulse, as our PSD variable. FIG.~\ref{fig:ej301_psdcalibration} shows F90 as a function of energy for $^{252}$Cf and $^{22}$Na sources.  Based on this calibration, neutron events will have F90 in the range 0.76-0.91, while gammas have faster pulses with F90 $>$0.91.  (Calculated values of F90 $>$ 1 may occur due to not accounting for baseline drift in our analysis.)  Discrimination with the PSD parameter degrades rapidly at low energies due to the limited photon statistics. PSD also fails in the high energy region above the digitizer saturation point, as described above. For this reason, we restrict the neutron analysis to the region between 50 and 700 p.e., which corresponds to approximately 0.3 to 1.6\,MeV imparted to the recoiling proton. 

\begin{figure}[t!]
  \centering
  \includegraphics[width=3.6in]{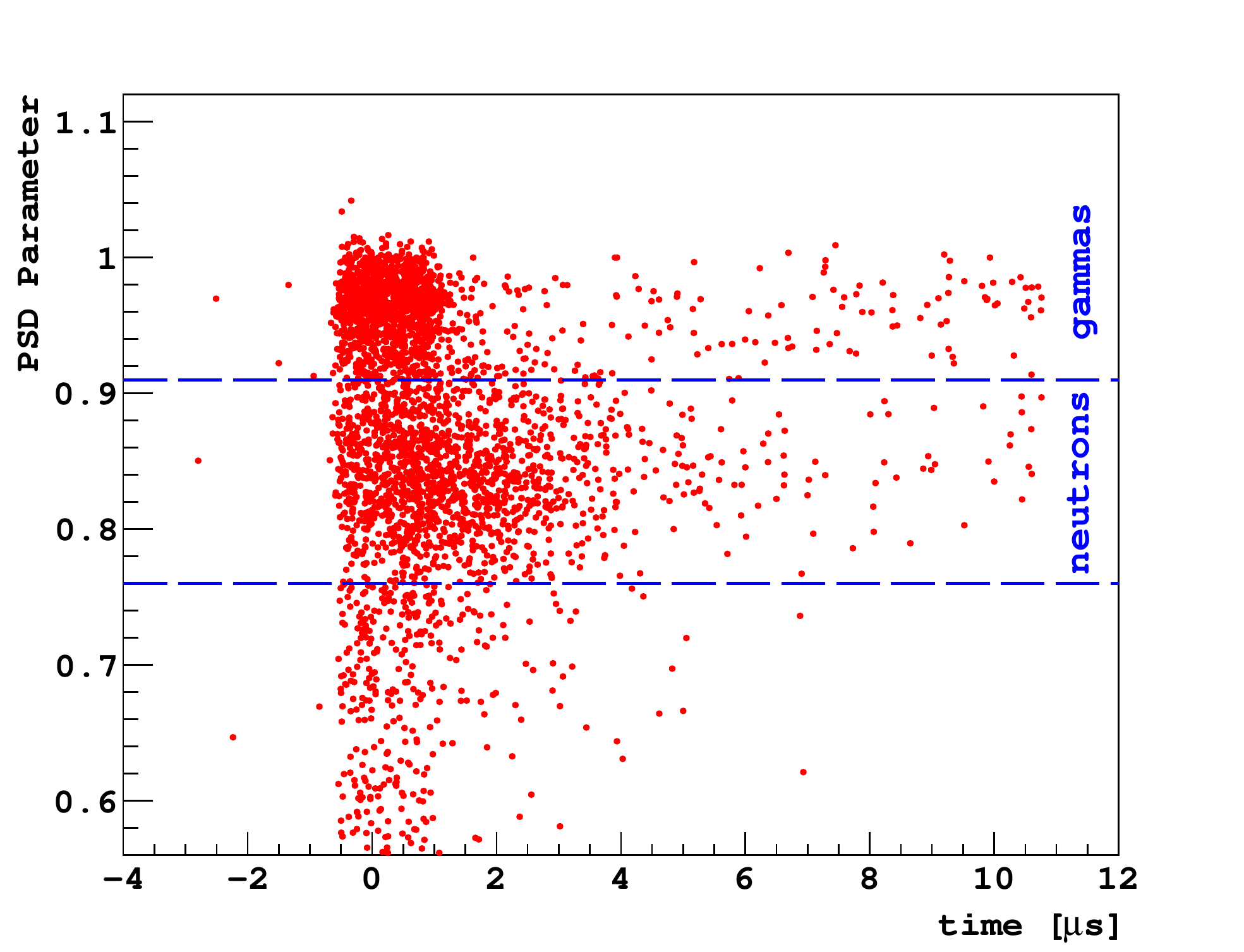}
  \includegraphics[width=3.5in]{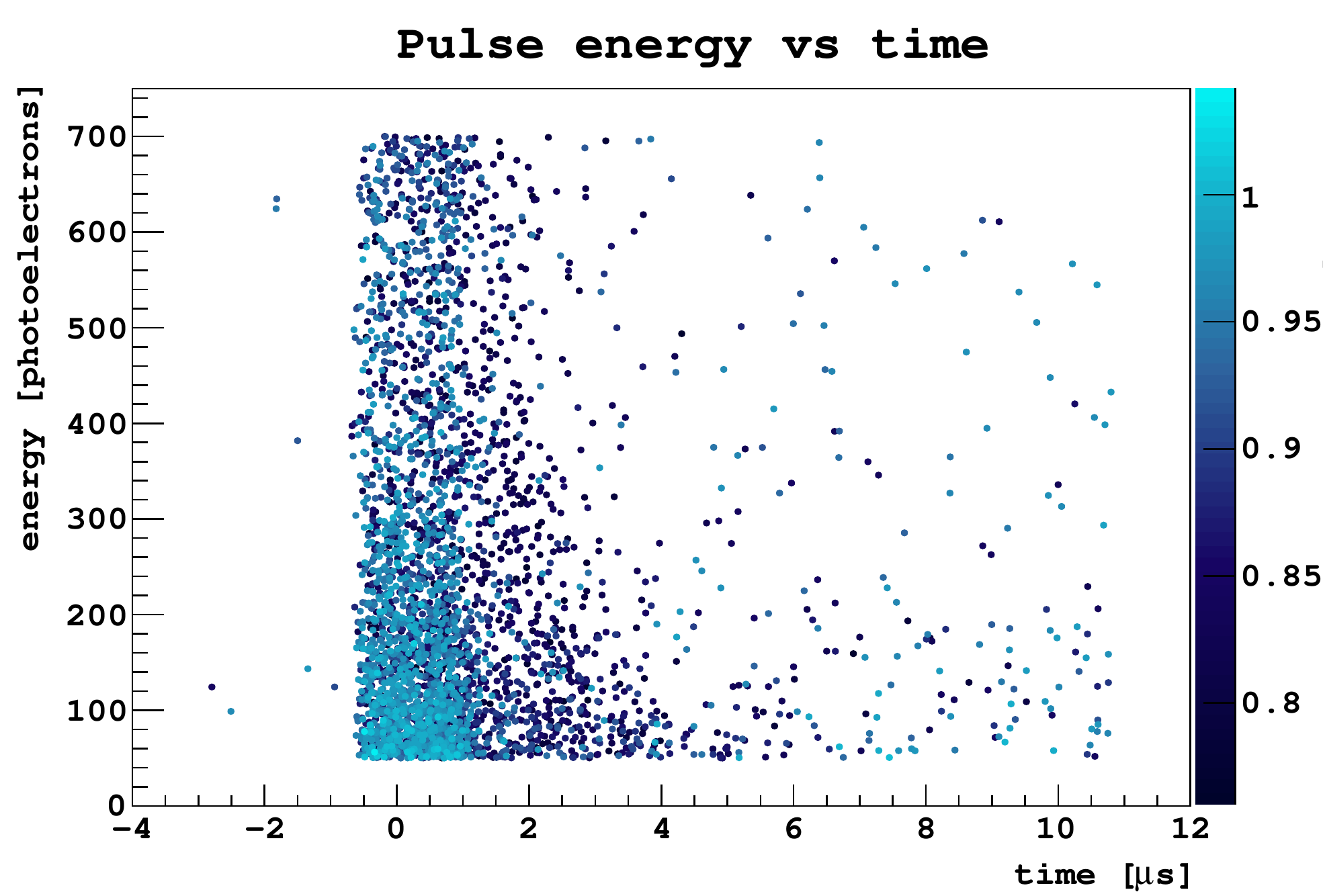}
  \caption{{Top: Pulse shape discrimination parameter of  scintillation events in the EJ301 detector vs time relative to the beam trigger. Bottom: Energy of scintillation pulses measured in the 5'' EJ301 detector vs time. The color scale for each point shows the PSD parameter, with darker colors being more neutron-like.}}
  \label{fig:ej301}
\end{figure}

FIG.~\ref{fig:ej301} top shows the F90 parameter vs detection time for events in the 50--700\,p.e. range. The tail of events with F90 $<$ 0.75 is most likely due to pileup events. The 1.7\,$\mu$s beam spill is evident in the region from -0.6 to 1.1\,$\mu$s on this time scale, and the events in this region are overwhelmingly gamma-like; after the spill, the rate is dominated by neutron-like events. The rate of neutron-like events peaks partway through the beam spill, then decays away with a characteristic time of a few $\mu$s. FIG.~\ref{fig:ej301} (bottom) shows the event energy as a function of time, from which it is clear that the energy of the neutron-like events also decays with the same few $\mu$s timescale. Both of these observations are roughly what one would expect from neutrons gradually losing energy to elastic scattering in the shielding material and the building. 

%FIG==============
%FIG==============

Although this measurement lacked precise calibration and required a small analysis window, with this first look we were able to determine the overall scale of the neutron background. Very few cuts were placed on the analysis, but those that were applied should have had the effect only of rejecting nuclear-recoil events while admitting a minimal amount of electron-recoil events. The measured rate of recoil-like events in the 0.3--1.6\,MeV range (assuming protons) in the liquid scintillator detector is $\sim$0.09 events per beam trigger. The average neutron-proton elastic scattering cross section in this energy range is approximately 6\,barns, which, given the 1.4~kg total scintillator mass, gives an average flux of $>$2$\times10^{-4}$/cm$^2$/pulse (pulse$\simeq$4.5$\times$10$^{12}$\,POT) neutrons with energy above 0.3\,MeV at 19\,m from the target. A more in-depth characterization of the beam-induced neutron flux requires a detector with larger mass, more dynamic range, and better particle discrimination, such as the SciBath detector.

%=====================================================================================
\subsection{SciBath detector}
%=====================================================================================
The SciBath detector is a prototype for the proposed FINeSSE detector which is a 13\,ton, fine-grained, liquid scintillator neutrino tracking detector~\cite{Bugel:2004yk}. While the detection concept was originally optimized to be a fine-grained neutrino tracker, it is also an excellent neutron detector. Below, we show results from a 2-month measurement of the beam-correlated neutron flux (10-200\,MeV) at the BNB target building. The SciBath detector will be described briefly here. More details about the SciBath detector will appear in a future publication.  

%-------------------------------------------------------------------------------------
\subsubsection{Detector Description}\label{sec:SBDetector}
\begin{figure}[t!] 
   \centering
   \includegraphics[width=3.in]{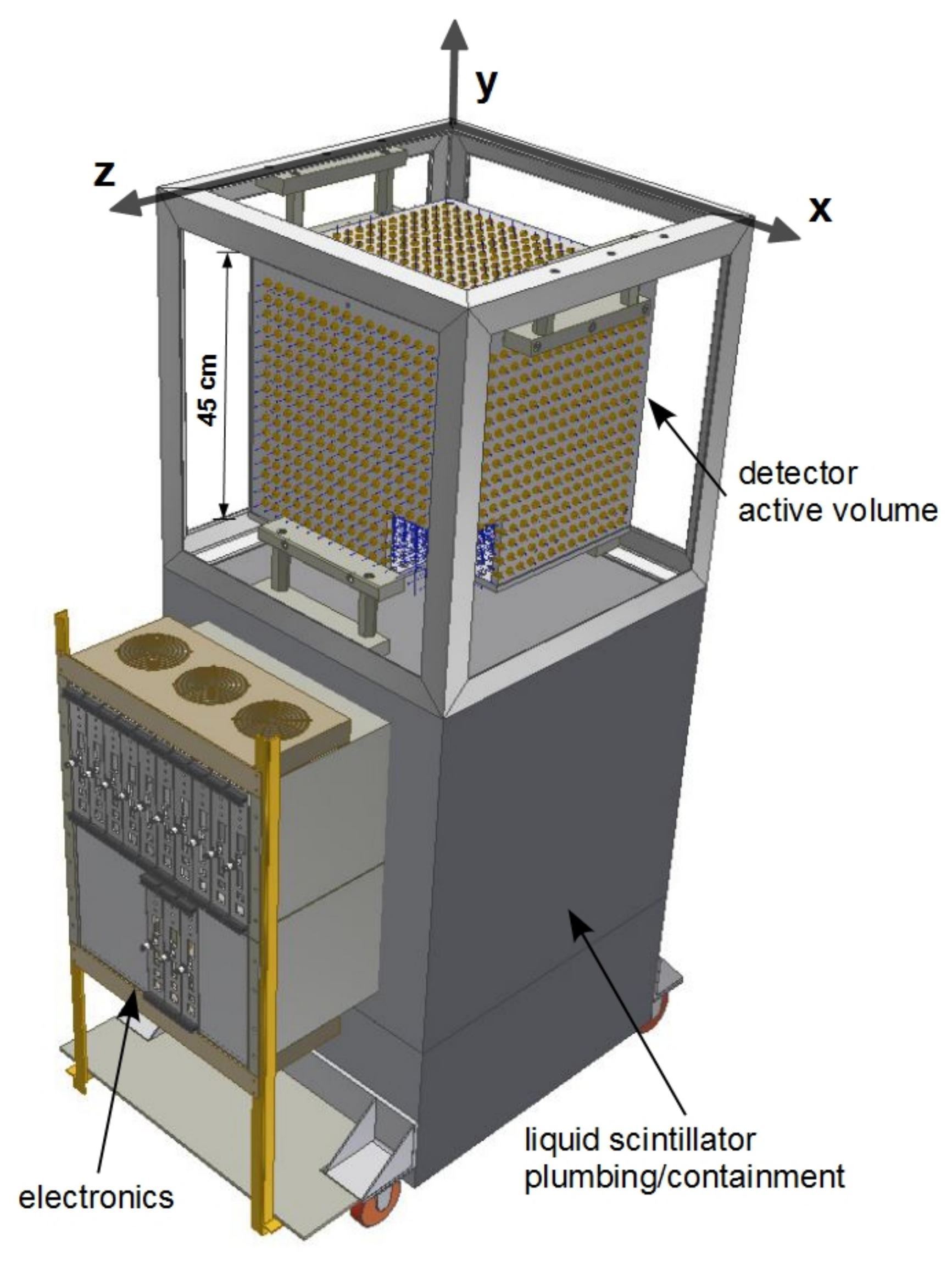} 
   \caption{{A schematic drawing of the SciBath detector with its (45 cm)$^3$ active volume indicated along with the other major components.}}
   \label{fig:SBapparatus}
\end{figure}

%-------------------------------------------------------------------------------------
The SciBath detector is an 82~L, optically-open bath of mineral-oil-based liquid scintillator that serves as both an active target and scintillator. Scintillation light is produced by the recoiling charged particles from neutral particle collisions with the mineral oil or by incoming charged particles from outside the detector. This scintillation light is absorbed by a square $16\times16$~array of wavelength shifting (WLS) fibers, oriented along each detector axis, with a spacing of 2.5\,cm (i.e.~768 total fibers). The light entering each fiber is Stokes-shifted and re-emitted isotropically. Some of the wavelength-shifted light is then transported by total internal reflection to a multi-anode PMT where it is read out and digitized by the DAQ. WLS fibers shift the ultraviolet bulk scintillation light to blue where it more effectively couples to the PMT quantum efficiency peak. A schematic of the detector is shown in FIG.~\ref{fig:SBapparatus}.  

%FIG==============
%FIG==============

The liquid scintillator has a base of mineral oil combined with 15\% pseudocumene (1,2,4-trimethylbenzene, C$_9$H$_{12}$) by volume and 1.5\,g\,/\,L PPO (2,5-diphenyloxazole, C$_{15}$H$_{11}$NO). The mixture was created for this detector and was continuously purged with N$_2$. It is very similar in composition to commercially available liquid scintillators EJ-321L~\cite{Eljen} and BC-517 H~\cite{StGobain}, but it lacks tertiary wavelength shifters such as bis-MSB or POPOP. The scintillator emission peaks at approximately 370\,nm, and the attenuation length for this light is over 1\,m in the detector and is adequate for the WLS fiber spacing. The 1.5\,mm diameter WLS fibers have an absorption peak at 345\,nm, and reemission peaks at 435\,nm which matches the peak quantum efficiency of the PMT. Approximately 8\% of this reemitted light in the WLS fiber is collected at the PMT.

The SciBath optical properties were calibrated with cosmic ray muons and an LED pulser system. A minimum ionizing muon will deposit approximately 65\,MeV into the SciBath detector and this yields approximately 400 detected p.e.. The energy deposit to light output is 6~p.e\,/\,MeV, and we found this calibration to be stable to within 5\% over the entire 2\,month run. Birks' law is used to model quenching effects for large $dE\,/\,dx$ particles (e.g.~protons). The Birks' law coefficient $kB$ used in the Monte Carlo simulation is 0.013\,g\,cm$^{-2}$\,MeV$^{-1}$ while KamLAND reports $0.0092 \pm 0.0001$\,g\,cm$^{-2}$\,MeV$^{-1}$ for the commercially similar BC-517H~\cite{Braizinha:2010zz}. A pulsed LED system was coupled to the opposite end, with respect to the PMT, of each WLS fiber. Low-light LED pulses were used to measure the single p.e. response of the PMTs and calibrate the SciBath DAQ. These LED calibrations were performed every three weeks, and the gains were stable to within 10\% throughout the entire run. In fact, they were stable when compared to a previous deployment six months prior.  

Each PMT is mounted to a custom, Indiana-designed, ``Integrated Readout Module'' (IRM) which serves as both a digitizing readout and physical mounting for the PMT. They are built on a VME form factor, but they are externally powered and connectivity is established through 1-Gigabit ethernet (in lieu of the VME power and connectivity standard). The front-end electronics of the IRM shapes and stretches the incoming pulses to enable simultaneous nanosecond timing resolution and spectroscopy with 20~MS\,/\,s, 12-bit flash ADCs. Additional processing with onboard FPGAs and an ARM-9 microcontroller digitize and transfer 64 PMT channels simultaneously. For data collection, the DAQ was externally triggered on the beam for 20\,ms with a 1/3~p.e. threshold per channel and 100\,$\mu$s of pre-trigger data. The LED calibration runs were also externally triggered, but only recorded 1\,$\mu$s of data with no zero-suppressing threshold.     

To exploit the tracking capabilities of 768 WLS fibers for a large number of events, fast algorithms were developed to determine the track-like properties of each event. The first four statistical moments of the WLS fiber light output are calculated for each axis.  
A principal component analysis is then performed giving characteristic eigenvalues and eigenvectors of the fiber hit distribution. 
Point-and track-like objects can be discriminated by their characteristic eigenvalue spectra.  
Additionally, a pair of likelihoods are created to further discriminate point- and track-like events.  
In this analysis, event topology is not used to construct the beam-correlated neutron spectrum, but it is used as a quality cut to select track-like events for the direction spectrum.   

%-------------------------------------------------------------------------------------
\subsubsection{Results}\label{sec:SBResults}
%-------------------------------------------------------------------------------------
The SciBath detector was placed about 20\,m away from the BNB target at nearly $180^{\circ}$ with respect to the beam direction, and the detector position is shown schematically in FIG.~\ref{fig:mi12shielding}. SciBath recorded ten-minute beam-on data runs starting on February 29, 2012 and ending on May 3, 2012 with a 95\% total livetime. After the BNB shut down on April 23, 2012, various calibrations were performed. During the entire run, $4.90 \times 10^{19}$ protons on target (POT) were delivered to the BNB target. Approximately 5.5 weeks of production-quality data are used in the analysis below, and this data set contains $3.50 \times 10^{19}$ POT.  The remainder of the time was used for LED calibrations and other systematic checks. A total of 2.5~TB of data was collected with the majority of events having low fiber multiplicity ($<$5) these were unused in the analysis.

\begin{figure}[t!] %  figure placement: here, top, bottom, or page
   \centering
   \includegraphics[width=3.5in]{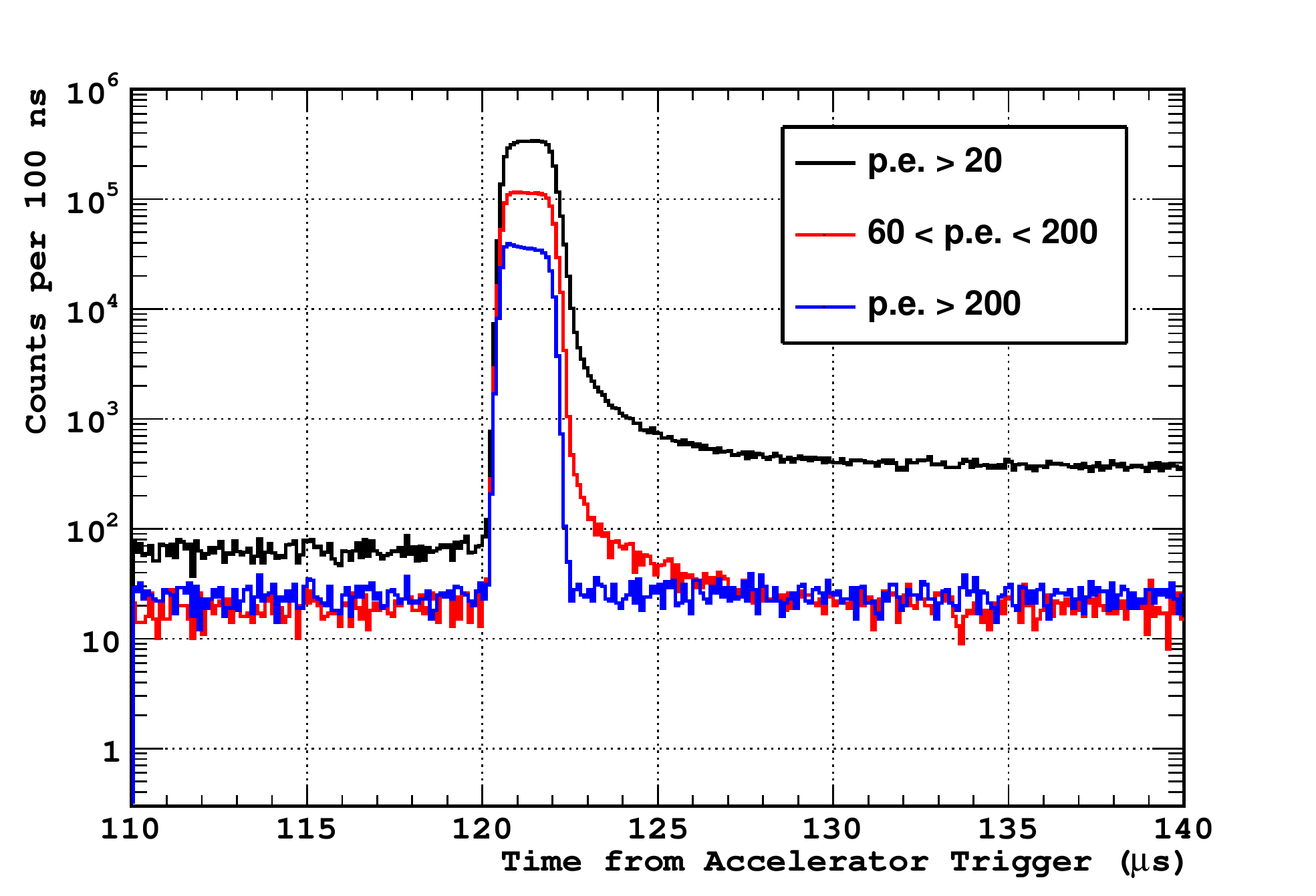} 
  \caption{{ The time distribution of events around the beam window for the given selection of p.e.. The dominant, black trace shows all events with p.e.~$> 20$, the red trace selects $60 <$~p.e.~$< 200$, and the blue trace selects p.e.~$> 200$. The red trace can be distinguished from the blue trace because it has an excess of events above background levels immediately after the beam is off. }}
   \label{fig:TimeDist}
\end{figure}

FIG.~\ref{fig:TimeDist} shows the distribution of events in time around the beam window for various p.e. subgroups.  
The black trace with the highest count rate is all events with p.e.\,$> 20$. The red colored trace is the group of events with $60 <$~p.e.~$< 200$ that has an excess of events above background for a few $\mu$s after the beam pulse. After which, the count rate returns to pre-beam, background levels. This is consistent with high-energy neutrons losing energy in the shielding, slowing down, and arriving at delayed times. On the other hand, the blue colored trace selecting events with p.e.~$> 200$ does not show an appreciable excess, and its count rate returns to background levels quickly after the beam pulse. The rate immediately after the beam for the p.e.~$> 20$ data remains significantly elevated above background levels for a longer timescale ($\sim 200~\mu$s).  This is consistent with the 2.2\,MeV, neutron-capture gamma rays from the hydrogen in mineral oil. For low event rates, neutron capture tagging can be used to discriminate primary neutrons from gamma rays, but this is not possible here because of the high event rate per beam spill. Correlating a specific neutron-capture candidate to a specific neutron primary scatter is impossible.

A minimal set of cuts is used to select events for analyzing the neutron energy spectrum and the high-energy neutron direction spectrum. A 3\,$\mu$s window surrounding the beam from 120\,$\mu$s to 123\,$\mu$s after the accelerator trigger is used to select in-beam events (see FIG.~\ref{fig:TimeDist}). Also, background events are selected in a 10\,ms window from 9\,ms to 19\,ms after the beam trigger and scaled for subtraction. For the neutron energy spectrum, events with p.e. $>$ 60 are selected to minimize the gamma-ray contamination, and for the direction spectrum events with p.e. $>$ 700 are selected to choose track-like events.

%FIG==============
%FIG==============

%-------------------------------------------------------------------------------------
\subsubsection{Direction Spectrum} \label{sec:Direction}
\begin{figure}[t!] %  figure placement: here, top, bottom, or page
   \centering
   \includegraphics[width=3.5in]{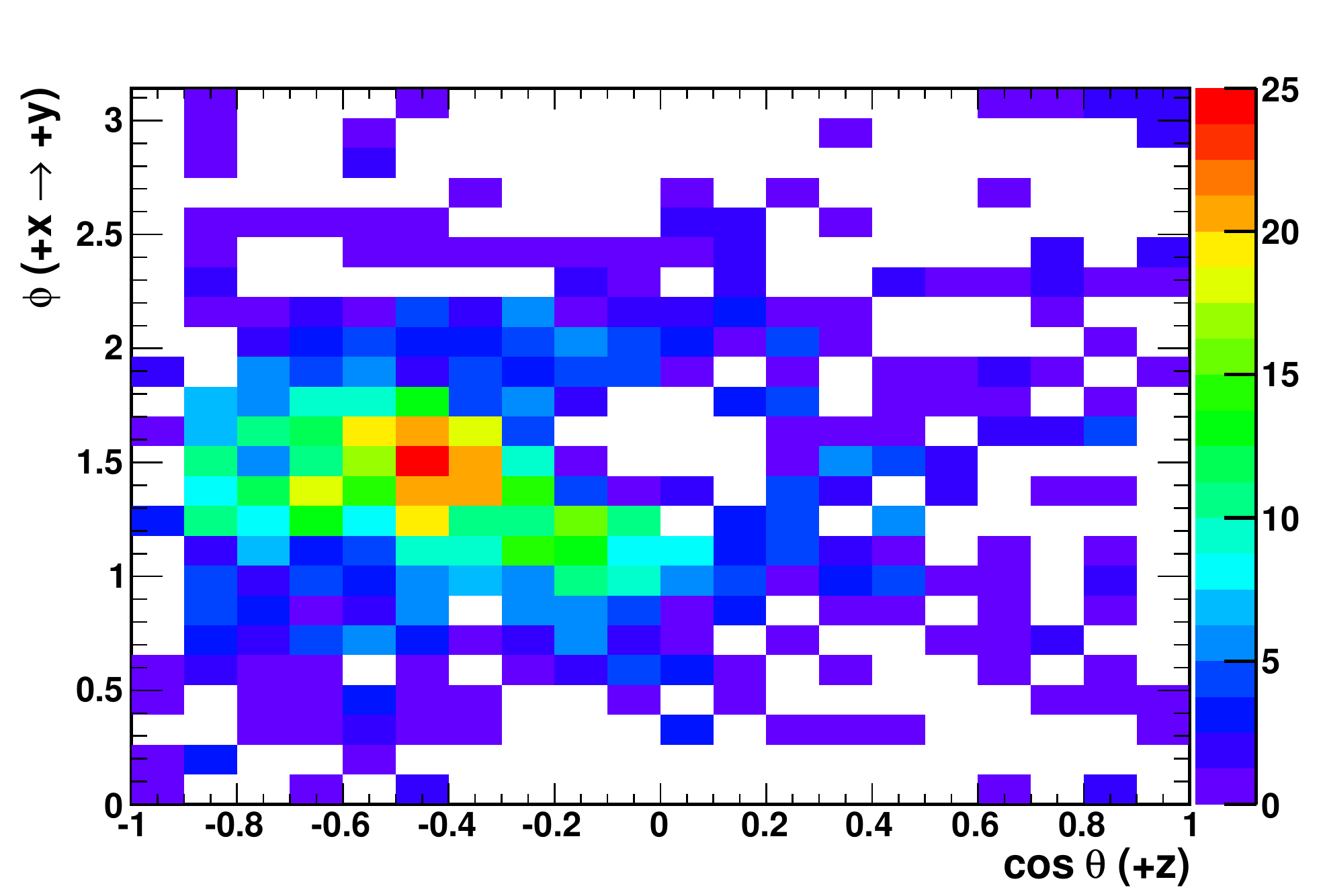} 
   \caption{{Direction spectrum for high energy proton recoils with track-like fiber hit distribution.  
   		In our right-handed coordinate system, $+z$ is pointed towards the beam target and $+y$ is vertical.  
		Back-projecting the peak of this distribution points in line with the beam, but about 10\,m upstream of the target.  }}
   \label{fig:BeamDir}
\end{figure}

%-------------------------------------------------------------------------------------
The direction-spectrum for high-energy proton recoils with track-like detector response is measured. 
In addition to the p.e. and timing cuts described above, events are required to be reconstructed within the inner 20\% fiducial volume and a modest set of track-like quality cuts are made.  
FIG.~\ref{fig:BeamDir} shows the proton recoil direction spectrum for energetic proton recoils after cuts. Back-projecting the peak of the direction spectrum locates a possible neutron source that is approximately 10\,m upstream of the BNB target. The spatial distribution of point-like events within the SciBath detector corroborates this result. The tracking capabilities were validated against the cosmic ray muon background and muon flux from the NuMI beam during a previous deployment.  
When validated against the cosmic ray spectrum, our results agree with the results of Mei and Hime~\cite{Mei:2005gm} and Miyake~\cite{Miyake:1973qk} to within 10\%.  

%FIG==============
%FIG==============

%-------------------------------------------------------------------------------------
\subsubsection{Neutron Energy Spectrum} \label{sec:En}
\begin{figure}[t!] %  figure placement: here, top, bottom, or page
   \centering
   \includegraphics[width=3.6in]{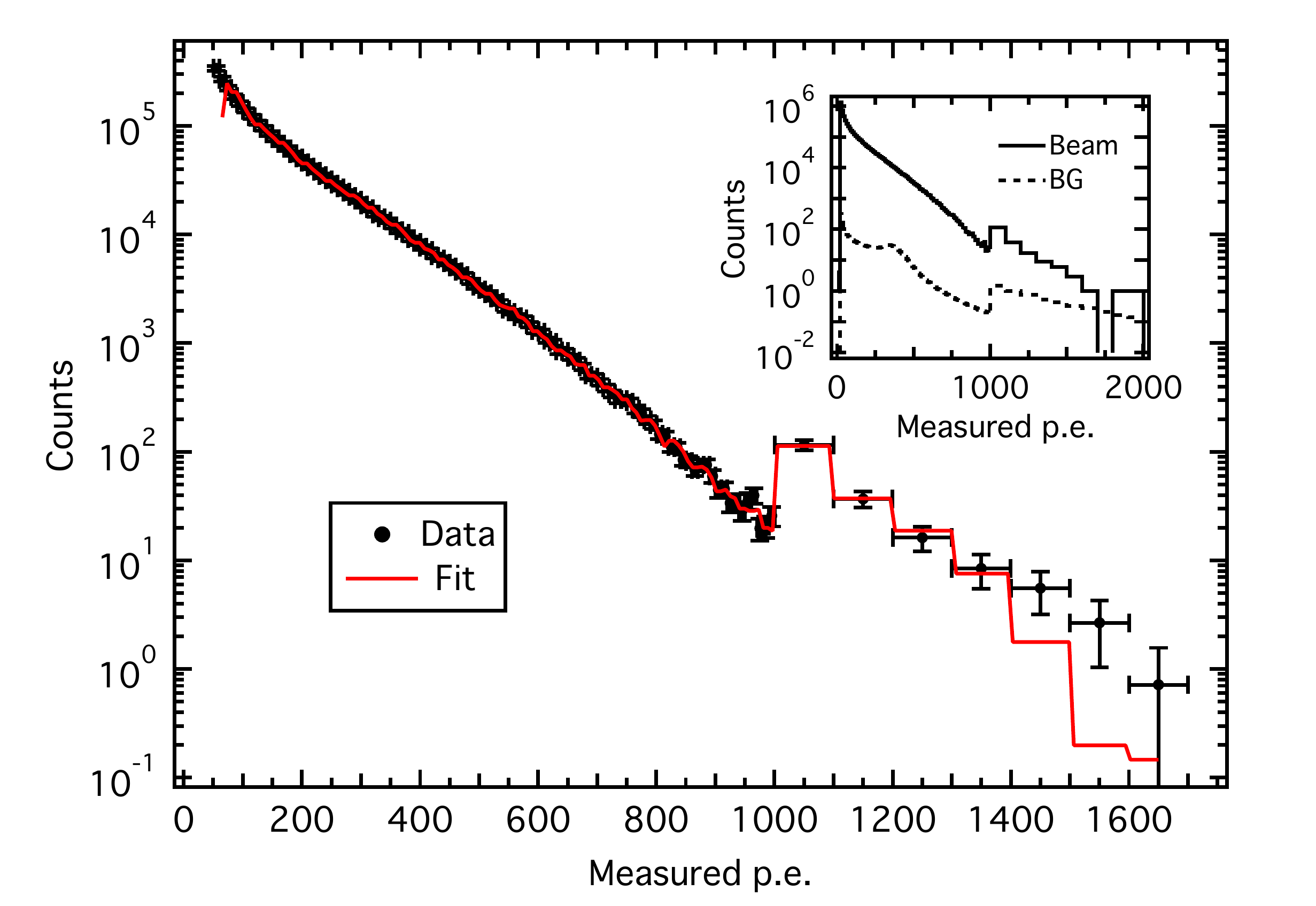} 
   \caption{{The p.e. spectrum used to unfold the neutron energy spectrum.  
   		Variable width bins are used with 10 p.e. bins below 1000 p.e. and 100 p.e. bins above. The insert figure shows a comparison between in-beam measurements (black line) and background measurements (dashed line: 9 to 19 msec off from the beam trigger).}}
   \label{fig:PEdata}
\end{figure}

%-------------------------------------------------------------------------------------
To analyze the neutron energy spectrum, the in-beam p.e. spectrum is background-subtracted for the entire data set.  
As shown in FIG.~\ref{fig:PEdata}, the in-beam rate clearly dominates the background rate when scaled for the total beam exposure time of 23\,s.  
The background subtracted data shows a cutoff at 1600~p.e. and this is consistent with the maximum SciBath response to a single, 200\,MeV proton recoil.
Higher p.e. events are occasionally observed, but their origin is consistent with hadronic cascades and multiple, energetic scattering events. 

%FIG==============
%FIG==============

The neutron energy spectrum is then unfolded from the p.e. spectrum by using the SciBath detector response as calculated with a Monte Carlo simulation (MC). From the results of the direction spectrum, we simulated a diverging beam of neutrons with a large cross-sectional area impinging on the SciBath detector. The simulation shows that SciBath has a 0.19\,m$^2$ effective cross-sectional area for neutron acceptance. Neutrons were uniformly generated up to 200\,MeV in 20\,MeV bins, and the simulation then tallied the total p.e. response for each 20\,MeV neutron energy bin. The p.e. response was binned in the same way as the data, 10~p.e. bins for p.e.~$< 1000$ and 100~p.e. bins for p.e.~$\geq 1000$. A least-squares fit was performed with each 20\,MeV neutron energy bin scaled by an independent fit parameter. During the fit, these fit parameters were constrained to be strictly decreasing as the neutron energy increases. These constraints were relaxed and other simulation configurations were tested in systematics tests. The resulting neutron energy spectra changed very little as the constraints were relaxed.  

\begin{figure}[t!] %  figure placement: here, top, bottom, or page
   \centering
   \includegraphics[width=3.5in]{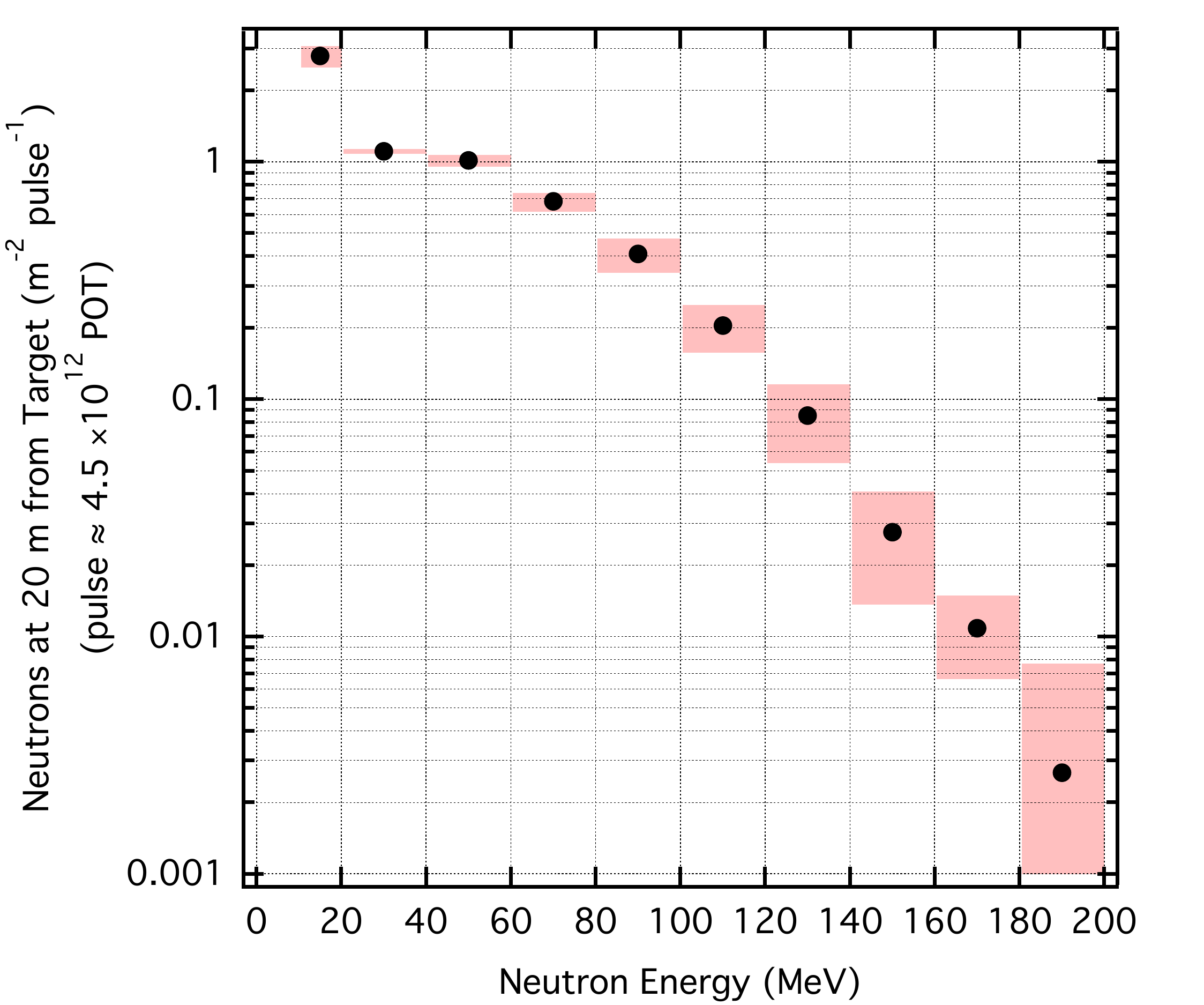} 
   \caption{{The measured neutron energy spectrum by SciBath 20\,m behind the proton target is shown.  
   		We measure $3.55 \pm 0.38$ neutrons per m$^2$ per beam pulse above 40\,MeV, and the low energy bin is strongly influenced by the detector threshold. The SciBath sensitivity above 200\,MeV is significantly reduced, and these energy bins have large uncertainties.}}
   \label{fig:CENNS_En_preliminary_log}
\end{figure}

FIG.~\ref{fig:CENNS_En_preliminary_log} shows the unfolded neutron spectrum per pulse per m$^2$ with the systematic uncertainties added in quadrature with the fit uncertainty. 
The total energy resolution is approximately 30\% near the 60~p.e. threshold, and this gives an effective neutron energy threshold of approximately 10\,MeV.  
From the unfolded neutron energy spectrum, we find the total number of neutrons above 10\,MeV per pulse per m$^2$ is $6.3 \pm 0.7$. Shielding the low energy neutron flux should not be challenging, but shielding will moderate high energy neutrons to potentially dangerous energies in the CENNS detector. With this in mind, the neutron flux above 40\,MeV is particularly dangerous as a background, and we measure $2.4 \pm 0.3$ neutrons per pulse per m$^2$ above 40\,MeV. Above 200\,MeV, the SciBath detector loses sensitivity because recoiling protons at these energies are no longer fully contained by the detector. Fits above 200\,MeV show very little significance, and the correlation matrix for the fit shows that we are unable to differentiate higher energy neutrons from 200\,MeV neutrons.  

%FIG==============
%FIG==============

%-------------------------------------------------------------------------------------
\subsubsection{Systematic Uncertainties} \label{sec:Systematics}
%-------------------------------------------------------------------------------------
In the analysis, we identified four classes of uncertainties to the neutron energy spectrum: energy scale calibration, fiducial cut, fit uncertainty, and the threshold. The dominant uncertainty above 60\,MeV is due to extrapolating the energy scale calibration defined at approximately 400 p.e. (6~p.e.\,/\,MeV) from cosmic ray muons to higher energies. We found that this conversion factor varied by 5\% for a number of reasons: uncertainty of the muon path lengths, detector energy resolution, p.e. counting statistics, light collection efficiency as a function of position, muon input into the MC, and analysis cuts.  
Above the 10\,MeV neutron energy threshold, the variation of the Birks law coefficient $kB$ had a negligible impact when compared to the other systematic uncertainties.

At low neutron energy, the choice of fiducial cut, uncertainty of the p.e. threshold, and the fit contribute roughly equally to the total uncertainty. The extraction of the neutron energy spectrum with the unfolding procedure should be independent of the choice of the central detector fiducial if the MC is correct. The neutron energy spectrum in FIG.~\ref{fig:CENNS_En_preliminary_log} uses the entire detector, but we found very little variation even down to 10\% of the total volume. Its effect on low neutron versus high neutron energies can be understood, because attenuation at the detector edges is stronger for low energy neutrons, whereas high energy neutrons are more penetrating produce longer track proton recoils which with average positions closer to the center of the detector. Because we do not have neutron-gamma discrimination at low energies, we set the p.e. threshold to 60 to remove gamma rays below 10\,MeV. Due to gain shifts during the run and the extrapolation of the energy calibration to low energy, we found that a 10\% variation in threshold was reasonable, and we used the MC to examine this variation on the unfolded neutron spectrum. As expected, the threshold will vary the first bin (10-20\,MeV) very strongly, but has no effect above 40\,MeV.

%-------------------------------------------------------------------------------------
\subsubsection{Cosmogenic Neutron Flux}
%-------------------------------------------------------------------------------------
%FIG==============
\begin{figure}[htbp] %  figure placement: here, top, bottom, or page
   \centering
   \includegraphics[width=3.5in]{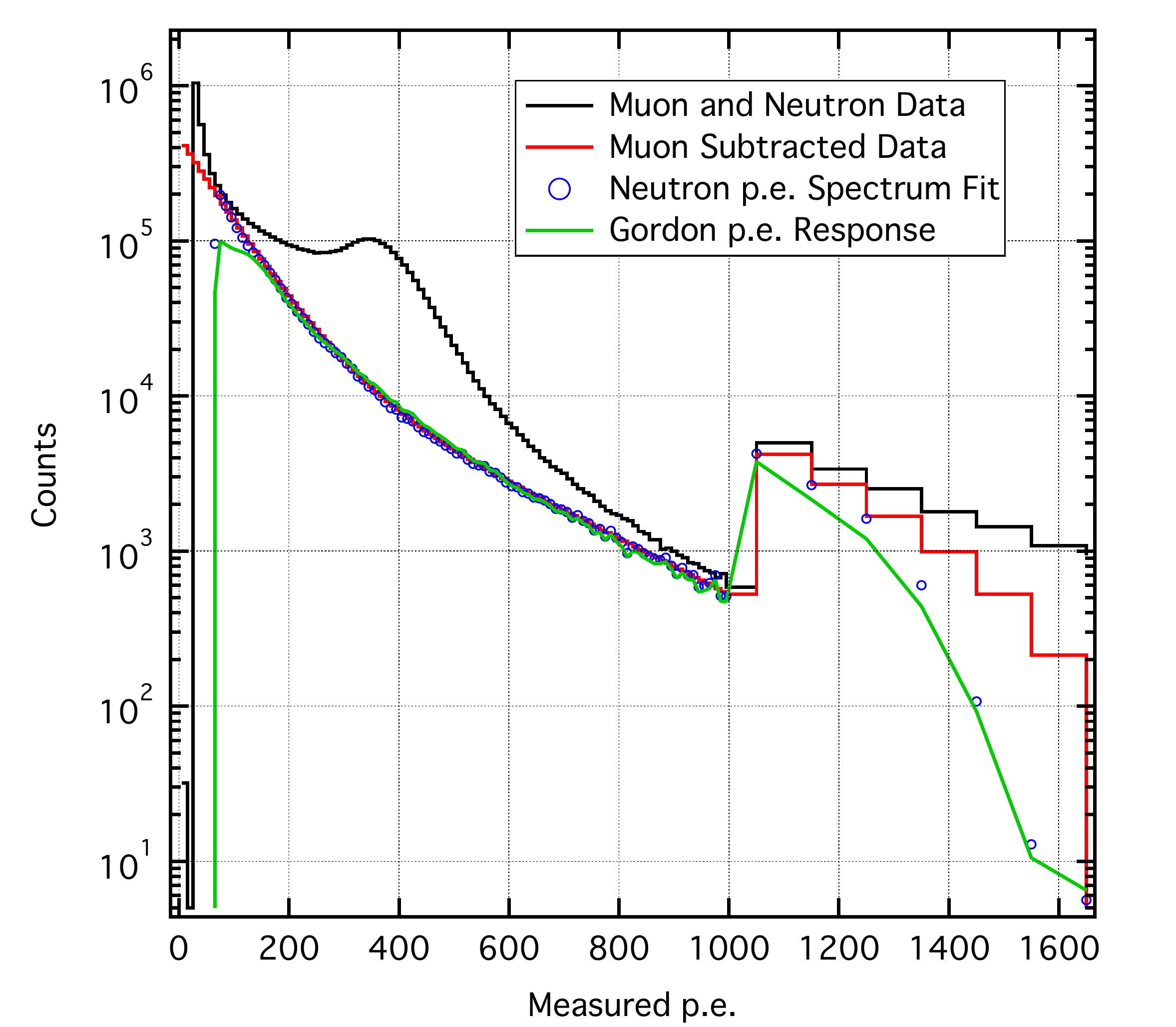} 
   \caption{{Raw cosmogenic p.e. spectrum (black) with double exponential fit to neutron data (red).  
   The unfolding process fits to the double exponential (blue markers), and the expected Gordon p.e. spectrum is also overlaid (green).  }}
   \label{fig:BG_PEs_fid8}
\end{figure}
%FIG==============
%FIG==============
\begin{figure}[htbp] %  figure placement: here, top, bottom, or page
   \centering
   \includegraphics[width=3.5in]{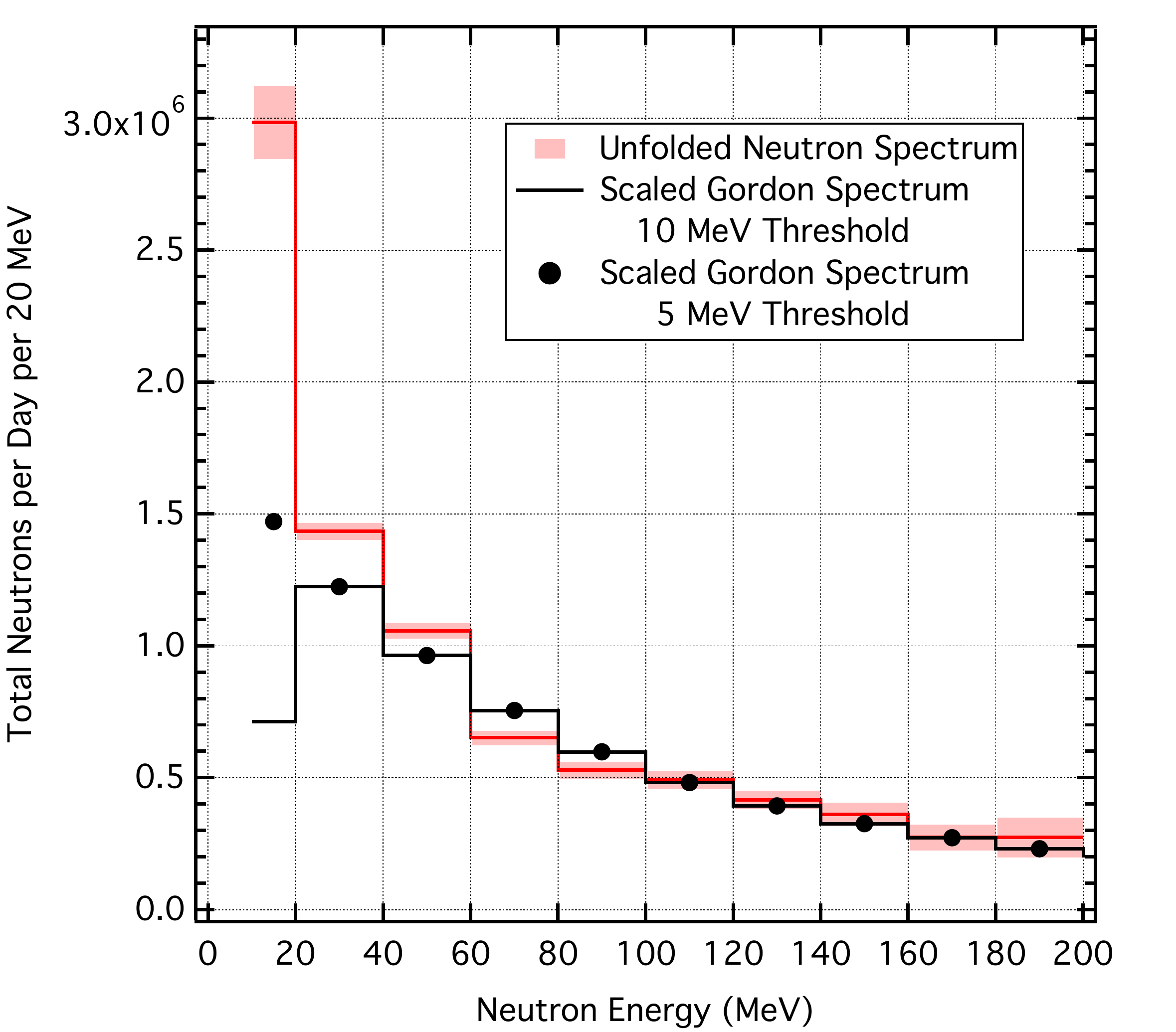} 
   \caption{{The unfolded neutron spectrum (red) overlaid with the Gordon neutron spectrum with 10\,MeV neutron energy threshold (black trace) and 5\,MeV threshold (black) markers.  }}
   \label{fig:BG_En_Spectrum}
\end{figure}
%FIG==============
For 10\,ms after each beam trigger, we collected background events with a total exposure of $8.5 \times 10^4$~s. 
The raw p.e. spectrum is shown in FIG.~\ref{fig:BG_PEs_fid8}, and the peak centered at 400 p.e. contains the minimum-ionizing, cosmic ray muons. To extract the neutron p.e. spectrum, the total p.e. plot is fit to a double exponential plus the muon response functions as calculated by the Monte Carlo. The double exponential is then fit with the same least-squares fitting procedure that was used for the in-beam data set. Gordon et al.~\cite{Gordon:2004} give a parameterization of the expected background neutron flux. For comparison, we applied our MC response function to the Gordon spectrum to generate the expected p.e. spectrum we would measure in our detector. The p.e. spectrum from Gordon was scaled by the effective area for neutron acceptance and by the total exposure time. To match the measured data, and additional scale factor 2.4 was required. The overlaid p.e. spectra are shown in FIG.~\ref{fig:BG_PEs_fid8}. FIG.~\ref{fig:BG_En_Spectrum} shows the unfolded neutron energy spectrum and the expected neutron spectrum from Gordon. Again, the Gordon spectrum requires a factor of 2.4 to match the neutron spectrum unfolded from our data. Aside from the factor of 2.4, our raw p.e. and unfolded neutron energy spectra shapes agree well with the parameterizations from Gordon above 20\,MeV. Our disagreement in the lowest energy bin seems to be indicative of threshold effects. The uncertainties shown are from the fit only, but the systematic uncertainty is similar to those in the in-beam data. \par

%%%%%%%%%%%%%%%%%%%%%%%%%%%%%%%%%%%%%%%%%%%%%%%%%%%%%%%%%%%%%%%%%%%%%%%%%%%%%%%%%%%%%%%
\section{CENNS Experiment}\label{cennsexperiment}
%%%%%%%%%%%%%%%%%%%%%%%%%%%%%%%%%%%%%%%%%%%%%%%%%%%%%%%%%%%%%%%%%%%%%%%%%%%%%%%%%%%%%%%
%=====================================================================================
\subsection{CENNS Detector}\label{detector}
%=====================================================================================
%FIG==============
\begin{figure}[t!]
\centering
\includegraphics[width=2.8in]{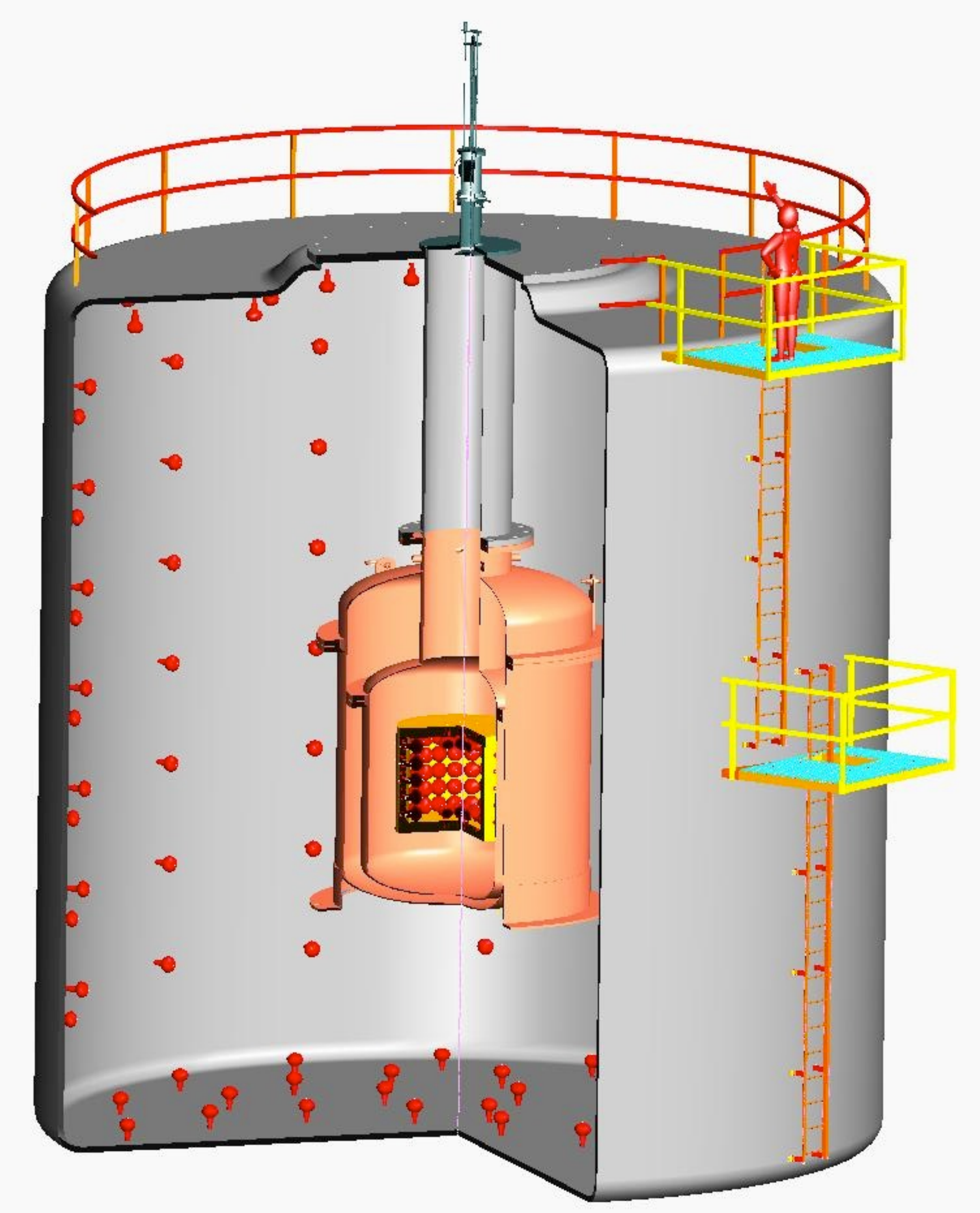}
{\caption{ Conceptual sketch of a ton-scale low energy threshold liquid argon detector. The active volume of the inner liquid argon detector is of a ton-scale and viewed by $\sim$100 low-temperature phototubes with 4$\pi$ coverage. The inner detector is enclosed in a vacuum insulation chamber. The outer water tank is designed for muon veto and neutron shielding.}\label{fig:detector}}
\end{figure}
%FIG==============

Liquid Argon (LAr) has several advantages as a detection medium. As in all of the noble liquids, LAr is naturally transparent to its own scintillation light and can be made very pure, leading to long attenuation lengths for the UV photons. Most critically, the time profile of the scintillation light created by the nuclear recoil signal is dramatically different than that for electron-like backgrounds. Radiation interacting with a noble liquid leads to the formation of dimers, in the form of trapped exciton states~\cite{PhysRevB.13.1649}. Both singlet and triplet states are formed and create ultraviolet scintillation light at 128\,nm when they decay. The lifetimes of these states are very different in LAr; 6\,ns for the singlet and 1.6\,$\mu$s for the triplet. Moreover, the relative amplitudes of these states depend on the type of ionizing radiation~\cite{PhysRevB.27.5279, Doke:1988dp, Doke2002}. Boulay and Hime~\cite{Boulay:2006mb} recognized that this Pulse-Shape Discrimination (PSD) allows for unprecedented rejection of $^{39}$Ar beta decay background intrinsic to the argon target, a concept that has since been demonstrated in small prototype detectors~\cite{Lippincott:2008ad, Benetti:2007cd} and has led to major efforts for the direct detection of dark matter.

Of particular utility to a CENNS measurement is the so-called ``single-phase'' approach to dark matter wherein only the primary scintillation light is recorded~\cite{Boulay:2006mb}. This approach allows one to design a detector with the high photo-coverage necessary to achieve the desired light yield and low-energy threshold. The PMTs are the only active component in the detector, affording simplicity in design. Moreover, the speed for recording digital pulses is governed by the triplet lifetime of the argon scintillation light, thus avoiding difficulties with pulse pileup and dead time associated with a time projection chamber.

The basic conceptual design of a single-phase detector is shown in FIG.~\ref{fig:detector} which is similar to the CLEAR detector concept~\cite{Scholberg:2009ha}. Key to measuring CENNS is a detector with a sufficiently large target mass and low-energy threshold to reveal a clean nuclear recoil signal that is free of background. The detector requirements for a CENNS measurement are similar to those for dark matter detection, however, with key differences: dark matter detectors need to be operated deep underground and free of cosmic ray induced background while a CENNS detector would be placed on the surface in a neutrino beam with its associated beam-related backgrounds. A great advantage of exploiting the BNB at Fermilab comes from its short-pulse time structure which provide a 5$\times$10$^{-5}$ reduction factor against steady state backgrounds.

%FIG==============
\begin{figure}[t!]
\centering
\includegraphics[width=3.5in]{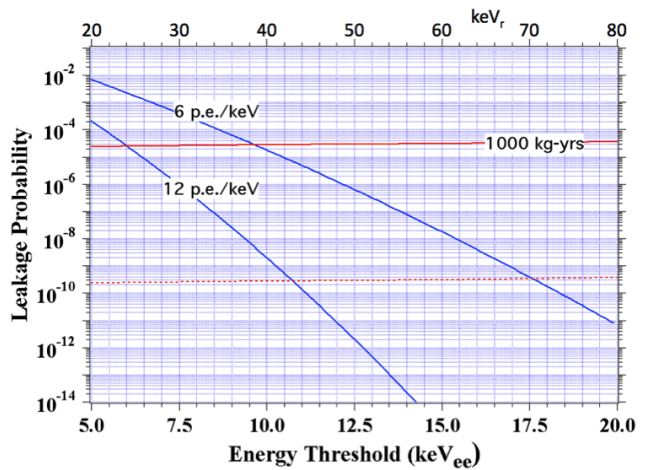}
{\caption{ Leakage probability of the internal $^{39}$Ar background as a function of energy threshold. The dotted red line indicates the statistical leakage rate of $^{39}$Ar events into the signal region for 1\,ton-year detector livetime. The solid red line indicates the leakage rate tolerable after duty factor correction ($5\times10^{-5}$). Solid blue curves show the impact of PSD cuts in the leakage probability for two different light yield assumptions.
 }\label{fig:leakageprob}}
\end{figure}
%FIG==============

%TAB==============
\begin{table*}[t!]
\begin{center}
	\begin{tabular}{  c | c | c  c  c  c  }
    \hline
    Energy Threshold  & Signal    &  Background    &        &	    & 	\\
	(keV$_{ee}$/keV$_{nr}$) &           & 6\,p.e./keV$_{ee}$ &   8\,p.e./keV$_{ee}$  & 10\,p.e./keV$_{ee}$   & 12\,p.e./keV$_{ee}$ \\  \hline
     5/20             &     320   & 228            &   69             & 21.6            & 6.8 \\ 
     7.5/30           &     196   & 11.5           &   1.5            & 0.21            & 0.03 \\ 
     10/40            &     136   & 0.45           &   0.02           & 0.001            & --- \\ \hline     
    \hline
    \end{tabular}  
    {\caption{ CENNS signal and $^{39}$Ar background (events/year) for a 1\,ton detector assuming 50\% acceptance in rejecting electron and gamma background. The background rate is determined for the energy window between energy threshold and 100\,keV$_{nr}$ (25\,keV$_{ee}$).}\label{tab:39ar}}
\end{center}
\end{table*}
%TAB==============

By far the largest activity in the detector arises from $^{39}$Ar in the LAr target. $^{39}$Ar is a beta emitter ($^{39}$Ar $\rightarrow ^{39}$K$ + e^{-} + \bar{\nu}_e$, Q = 535\,keV, $\tau_{1/2} = 269$\,year). In natural argon it is present at $\sim$8 parts in 10$^{16}$, yielding the decay rate of $\sim$1 Bq/kg. The unique attack on this intrinsic background is PSD. As can be seen in FIG.~\ref{fig:leakageprob}, the ability to reject the internal $^{39}$Ar background is a very strong function of the light yield, which in turn dictates what can be achieved as an analysis energy threshold. The pulsed structure of the BNB provides significant reduction in this background. If, for example, we assume 6 p.e./keV$_{ee}$ for light yield, then one can expect to achieve an energy threshold of 10 keV$_{ee}$ (40 keV$_{nr}$) with leakage of only one $^{39}$Ar event in an exposure of 1\,ton-year. The assumption of 6 p.e./keV$_{ee}$ is based on that measured in microCLEAN and projected for MiniCLEAN using the Hamamatsu R5912-02MOD PMTs submerged and operating cold in LAr~\cite{Lippincott:2008ad, Hime2011}.

Significant improvements are foreseen with PMT technologies that increase the efficiency of 19\% for the R5912-02MOD to $\sim$32\%. Simplification and optimization of the optical light-guides designed for MiniCLEAN could also improve light yield by as much as 30\%~\cite{Alexander201344}; hence it is quite reasonable to consider a single-phase, LAr detector with light yield as high as 12 p.e./keV$_{ee}$. As can be seen in Table~\ref{tab:39ar}, this would yield a detector with an energy threshold as low as 6 keV$_{ee}$ (24 keV$_{nr}$) that is essentially free of steady state and detector-related background.

In addition to $^{39}$Ar in the sensitive volume, there are external backgrounds arising from the detector construction materials themselves. Table~\ref{tab:surfacebackgrounds} contains a projection of the non-$^{39}$Ar backgrounds after scaling the MiniCLEAN backgrounds to a 1-tonne detector target and appropriate surface area~\cite{Hime2011}. Unlike a dark matter detector, the CENNS detector can employ the full target mass without fiducialization since the duty factor of the BNB is such as to make the steady backgrounds from neutron backgrounds and surface radon progeny negligible. Therefore, CENNS experiment does not require this extreme level of radon background control. Hence, we assume 100~/m$^2$/day or lower of modest level radon daughter decay rate in the energy region of interest which is reasonably achievable~\cite{Boulay:2009zw}.

%FIG==============
\begin{figure}[t!]
\centering
\includegraphics[width=3.5in]{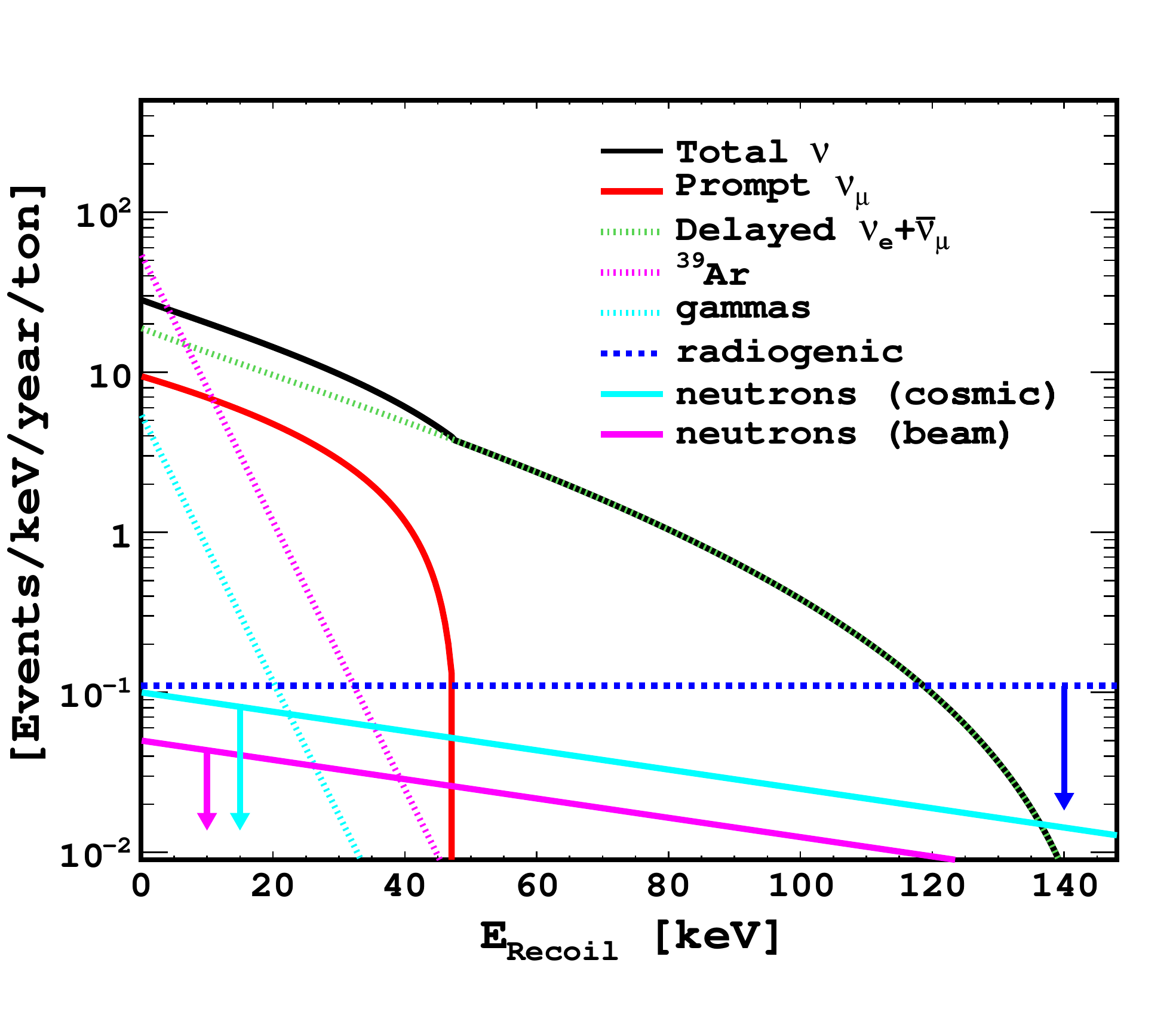}
{\caption{Number of expected CENNS events with far-off-axis BNB (32\,kW) neutrino flux. The liquid argon detector is assumed to be located at 20\,m away from the target. The beam-induced (cosmogenic) neutron background estimated based on SciBath measurements and assuming 7\,m (4\,m) of concrete shielding but without water shielding (see FIG.~\ref{fig:NeutronShielding}). Flat 50\% detection efficiencies are applied for nuclear recoil events.}\label{fig:evrate}}
\end{figure}
%FIG==============

%TAB==============
\begin{table*}[t!]
\begin{center}
	\begin{tabular}{  c | c  c  c  c  c  }
    \hline
Source           & Production Rate &  Detection Rate  & E $<$25 keV$_{ee}$  &	 12.5$<$E$<$25 keV$_{ee}$   	\\
                 &    (/ton/year)  & (events/ton/year)&                &                           \\  \hline
PMT($\alpha$,n)  &     66,700      &   11,340         &   1,520        & 710                       \\ 
Steel($\alpha$,n)&     3,680       &      495         &   65           &      30                   \\ \hline
Total($\alpha$,n)&     70,380      &     11,835       &   1,585        &      740                  \\  
Total($\alpha$,n)$\times$ duty factor & 3.5 & 0.6     &    0.08        &    0.04                   \\\hline\hline   
Radon            &    15,880       &                  &                   7,147 (25$<$E$<$100 keV$_{nr})$ \\\hline
Radon $\times$ duty factor &  0.8  &                  &                    0.36                  \\ \hline      
\end{tabular}
    
    {\caption{Backgrounds in the 1\,ton CENNS detector arising from ($\alpha$,n) neutrons from the PMTs and steel. The radon background is from TPB and acrylic.}\label{tab:surfacebackgrounds}}
\end{center}
\end{table*}
%TAB==============

	FIG.~\ref{fig:evrate} shows the event rate of CENNS in a one ton liquid argon neutrino detector given a neutrino flux of $5\times 10^5 \nu$/cm$^2$/s when the detector is located 20\,m away from the target at a far-off-axis site. Assuming flat $\sim$50\% detection efficiency, which is mostly from the PSD cut efficiency~\cite{Lippincott:2008ad, Benetti:2007cd}, we expect about $\sim$250 CENNS events/ton/year at 25\,keV$_{nr}$ energy threshold after background subtraction (at 32\,kW beam power). The beam-induced neutron backgrounds and systematic uncertainties are discussed in the following sections. \par

%-------------------------------------------------------------------------------------
\subsection{Neutron Shielding}\label{neutronshielding}
\begin{figure}[t!] %  figure placement: here, top, bottom, or page
   \centering
   \includegraphics[width=3.5in]{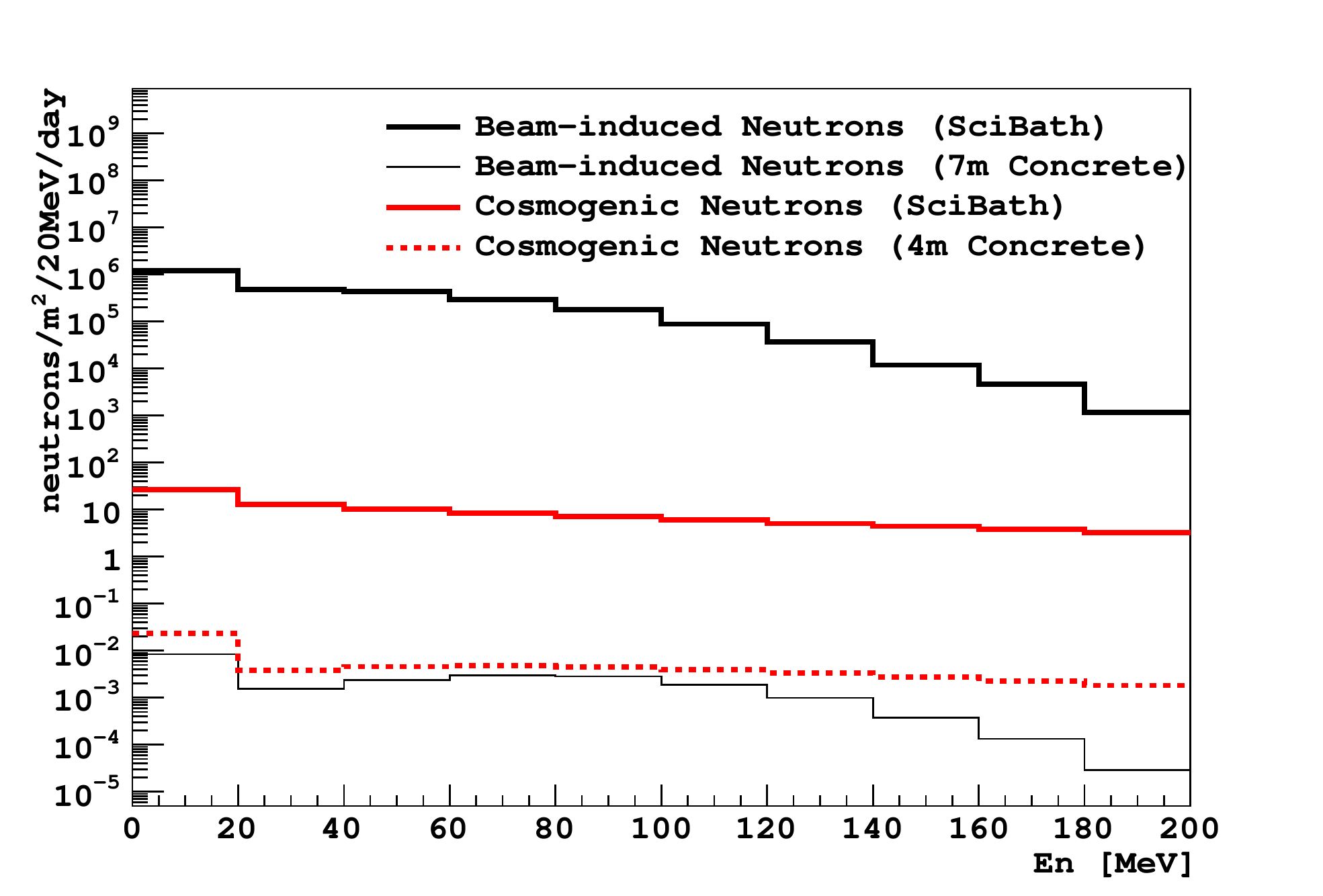} 
   \caption{{Neutron flux reduction with concrete shielding. The thick black solid line is beam-induced neutrons and thick red solid line is cosmogenic neutrons measured by SciBath detector. The detector livetime corrections are made for both input neutron fluxes. The thin black line is beam-induced neutron flux after passing 7\,m of concrete shielding. The dotted red line is cosmogenic neutron flux after passing 4\,m of concrete shielding. }}
   \label{fig:NeutronShielding}
\end{figure}

%-------------------------------------------------------------------------------------
The measured beam-induced neutrons (see FIG.~\ref{fig:CENNS_En_preliminary_log}) can be significantly reduced with proper shieldings. The fast neutron component, above 100\,MeV, requires special attention in shielding design. These neutrons may slow down in the shielding material itself and then become a more difficult background component with slower neutrons of less than a few MeV energy. We carried out MCNP and Geant-4 based Monte Carlo simulations in order to evaluate the overall level of neutron shielding that is needed for a CENNS experiment. We used the measured beam-induced neutron fluxes as input to the simulation. We found these neutron fluxes can be substantially suppressed by more than 7 orders of magnitude after 7\,m of concrete shielding. FIG.~\ref{fig:NeutronShielding} shows results of the MC from a Geant-4 based simulation. MCNP results are consistent with the Geant-4 results. We also found that measured cosmogenic neutrons can be significantly suppressed with 4\,m of concrete shielding. Given these levels of concrete shielding, the total number of neutrons that enter the detector's water shielding within the detector livetime can be less than 20 neutrons/m$^2$ per year of operation time. 

%FIG==============
%FIG==============

The neutrons entering the water shielding (10\,m in diameter) are then passed to the liquid argon detector in Geant-4 MC. In order to boost statistics of neutrons we simulated one million neutrons, then scaled to the expected input neutron fluxes. The resulting neutron-nucleus event rate in the liquid argon detector with water shielding is negligible (less than 10$^{-3}$ events/ton/year). Therefore in FIG.~\ref{fig:evrate}, we show MC results of neutron-nuclear recoil events without water shielding but with 7\,m concrete for beam-induced neutron shielding and 4\,m concrete for cosmogenic neutron shielding, in order to display the effect of our shielding option. To be conservative the neutron background events includes not only single scattering events but also multiple scattering nuclear recoil events in the LAr fiducial volume. The expected number of background events are only 1.4 events for beam-induced neutrons and 2.5 events for cosmogenic neutrons within the energy region of interest (25\,keV$_{nr}$ to 100\,keV$_{nr}$). This low-background configuration suggests that the CENNS detector can be placed as close as 14\,m away from the target where we expect twice the neutrino flux than at the 20\,m location. However it is also true that predicting the neutron flux over a massive shielding without accurate understanding of shielding configurations is quite challenging. Therefore, beam tests of various neutron shielding configuration would be needed. One important check is to see if the neutrons are from ``sky shine'', directly from the target or from the beamline.\par

%=====================================================================================
\subsection{Systematics  and Discovery Potential}\label{subsec:systematics}
%=====================================================================================
	There are two major sources of systematic uncertainties in a CENNS experiment: (1) Uncertainties of stopping pion production at the BNB target, and hence the related systematic uncertainties of absolute flux of neutrinos at the far-off-axis. (2) Uncertainties of scintillation yield ({\it Leff}) in liquid argon detector for the measurement of low-energy nuclear recoil events. The other sources of systematics such as beam-induced neutron backgrounds, cosmogenic neutrons, gamma backgrounds, ambient radioactive decays and uncertainties from high energy neutrino interactions near or in the detector, depend on the specific experimental design or are minor background contributions. 
	
\subsubsection{Uncertainty of neutrino flux} 
The uncertainty in neutrino production from stopped pions and muons is dominated by the uncertainty of the pion production in the BNB target and surrounding materials. The HARP experiment at CERN measured pion production from both thin beryllium targets and a replica BNB target at the 8\,GeV proton energy that the BNB uses. The uncertainty of the pion production measured by HARP was 7\%~\cite{Catanesi:2007ab,DSchmitz2008}. In addition to the uncertainty in direct pion production there are uncertainties that arise from the secondary production of pions and uncertainties in the fraction of pions and muons that get to decay rather than interact. These additional uncertainties are estimated to be at the 5\% level~\cite{AguilarArevalo:2008yp}. This gives a total of 9\% neutrino flux uncertainty.\par

%FIG==============
\begin{figure}[t!]
\centering
\includegraphics[width=3.5in]{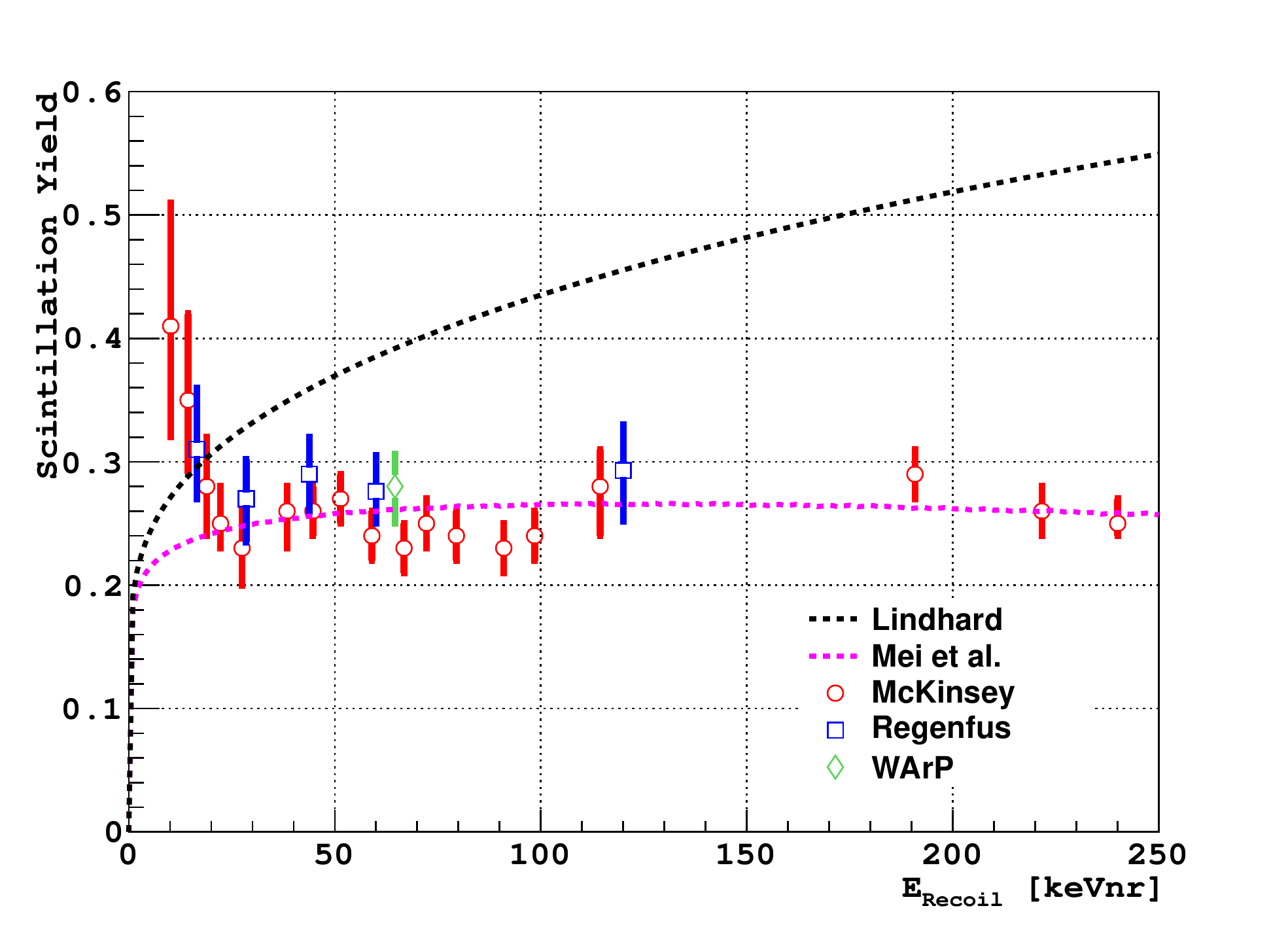}
{\caption{ The scintillation efficiency for nuclear recoils relative to the electron-equivalent measured in microCLEAN~\cite{Lippincott:2008ad}, Regenfus et al.~\cite{Regenfus:2006rt} and the single, averaged value from WArP~\cite{Brunetti:2004cf}. The model of Mei {\it et al.}~\cite{Mei200812} combines the Lindhard theory with BirkÕs saturation providing the phenomenological description indicated.}\label{fig:leff_vs_recoil}}
\end{figure}
%FIG==============
%FIG==============
\begin{figure}[t!]
\centering
\includegraphics[width=3.5in]{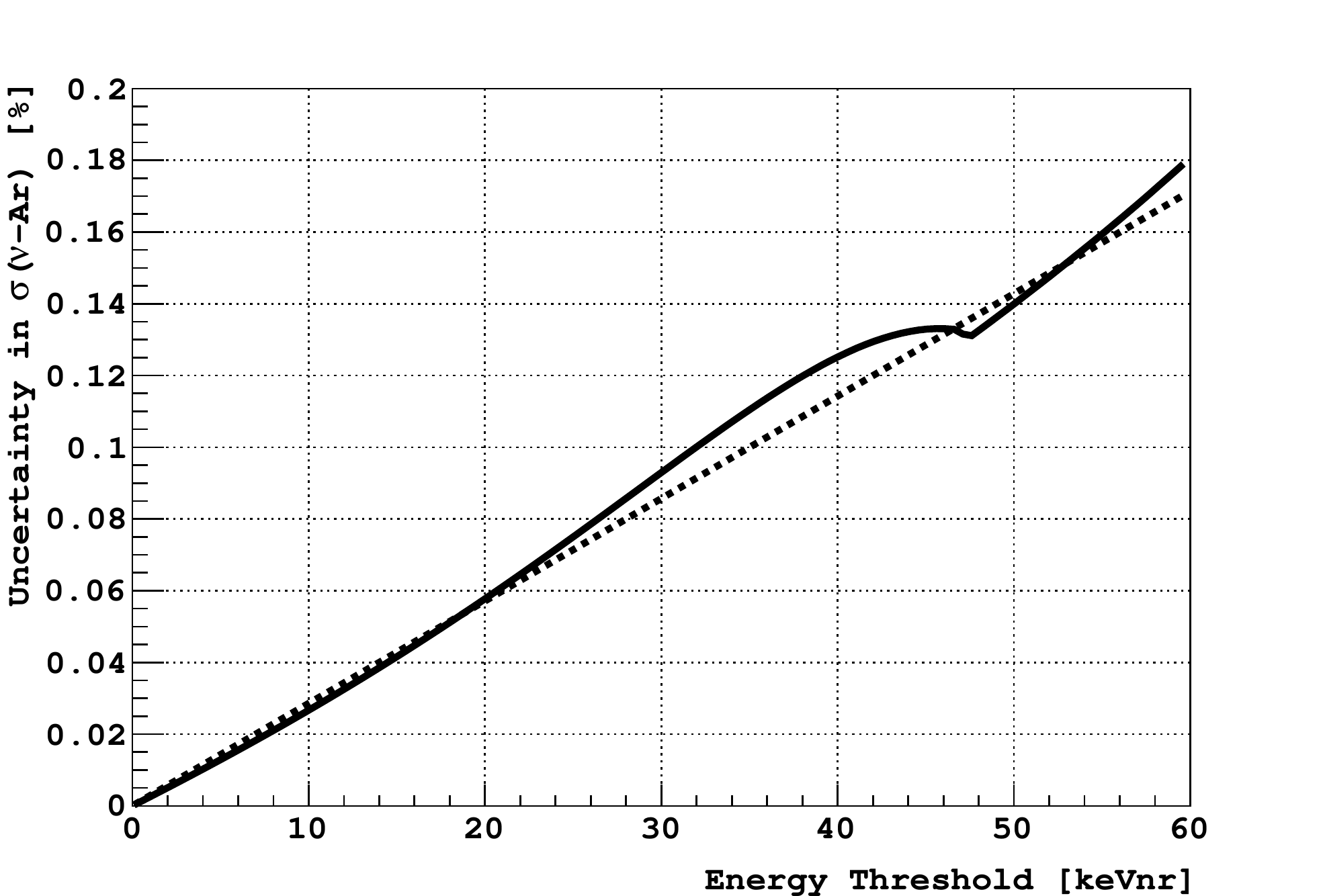}
{\caption{ Extracted cross section uncertainty as a function of energy threshold due to the intrinsic uncertainty in {\it Leff} of 6.5\%. The dashed curve indicates the uncertainty calculated using an analytical approximation to the shape of the differential neutrino-nucleus scattering spectrum and the solid curve uses the ÒtrueÓ spectrumÓ as simulated for the BNB. The uncertainty in {\it Leff} effectively induces an uncertainty in knowledge of the energy threshold and thus the integrated event rate above or below that threshold. The bump structure near $\sim$47\,keV$_{nr}$ comes from the similar structure in the event rate at the same energy (see FIG.~\ref{fig:evrate} black-curve) as the {\it Leff} is changing monotonically in these energies.}\label{fig:CSuncertaintyLeff}}
\end{figure}
%FIG==============

\subsubsection{Uncertainties from {\it Leff} of liquid argon detector} 
The scintillation efficiency for nuclear recoils relative to the electron-equivalent efficiency, referred to as {\it Leff}, has been measured for LAr in microCLEAN~\cite{Lippincott:2008ad} and independently by Regenfus et al.~\cite{Regenfus:2006rt} as 0.25$\pm$0.01$\pm$0.01 and 0.29$\pm$0.03, respectively. The data, shown in FIG.~\ref{fig:leff_vs_recoil}, are in good agreement with a model that combines simple Lindhard theory with BirkÕs saturation law~\cite{Mei200812}. The scintillation efficiency for nuclear recoils is essentially flat, independent of energy, for recoil energies above $\sim$20 keV$_{nr}$. The combined measurements provide {\it Leff} = 0.262$\pm$0.017. A possible up-turn in {\it Leff} at the lowest energies measured is interesting and worth further exploration. It is very likely due to an intrinsic energy dependence in the scintillation yield for gamma rays. Measurements are typically made of the nuclear recoil scintillation yield relative to a calibrated energy scale for gammas and it is usually assumed that the scintillation yield for gammas is independent of energy. FIG.~\ref{fig:CSuncertaintyLeff} shows expected cross section uncertainty as a function of energy threshold due to the {\it Leff}. At the energy threshold of 25\,keV$_{nr}$ the measurement uncertainty of cross section by {\it Leff} is 7.5\%. \par

\subsubsection{Uncertainties from high energy neutrino interactions}
The high energy neutrinos ($>$55\,MeV, see FIG.~\ref{fig:bnbangspc}-(b)) are produced by muon-capture and kaon decay at rest. These neutrinos represent only 3\% of the total neutrino fluxes.  However these high energy neutrinos may produce two types of background events; (1) direct neutrino interactions in the liquid argon volume, and (2) neutrino interactions in the water shield which result in secondary neutrons reaching the sensitive detector volume and leave nuclear recoils in the signal region. \par

We carried out a detector simulation for high energy neutrino interactions using FLUKA~\cite{fluka1,fluka2,fluka3}.  The neutrino interactions were weighted by neutrino-nucleus cross sections obtained with the GENIE(2.8.0)~\cite{genie} neutrino simulation package. Table~\ref{henbRates} shows the above two background cases. \par

%TAB============== 
\begin{table}[hbt]
   \centering {\small \begin{tabular}{l c c c c c}
    \hline & \multicolumn{2}{c}{\small Liquid argon} & \multicolumn{1}{c}{} &
    \multicolumn{2}{c}{\small Water shield} \\
                  & All events & w/neutrons  &  & All events & w/neutrons \\
    \hline \hline $\nu_{e}$ &    0.00 &    0.00 & &    0.00 &    0.00 \\
    $\nu_{\mu}$             &    0.39 &    0.28 & &    1.04 &    0.12 \\ 
    $\overline{\nu}_{\mu}$  &    0.04 &    0.02 & &    0.01 &    0.00 \\ \hline
        \hline
    {\bf Sum}               &    0.43 &{\bf 0.30}&&    1.05 & {\bf 0.12} \\ \hline
   \end{tabular} } \caption{Expected background events by E$_{\nu}>$\,55 MeV which deposit energy of 25\,keV to 100\,keV per ton liquid argon detector per year. The numbers of events with secondaries produced in (or reaching) the sensitive volume are presented in the `all events' columns. More critical events containing one or more neutrons are given by the `w/neutrons' columns.} \label{henbRates}
  \end{table}
%TAB==============

The CENNS signal is identified by single nuclear recoils in the energy range 25\,keV to 100\,keV, and the most serious background is expected from nuclear recoils caused by undetected neutron scattering. An upper limit of 0.42 events (=0.30+0.12 events or 0.21 events after applying 50\% detection efficiency) per ton-year is found for the neutrino-induced background. As the number of expected background events is small, statistical uncertainties in the simulation are not expected to be relevant. The largest systematic uncertainty of this study arises from the neutrino-argon cross-section uncertainties in the GENIE model in the relevant neutrino energy range (55\,MeV to 250\,MeV), which has never been measured. However, even if we assume an order of magnitude of uncertainty in the GENIE cross section model in this energy region, the backgrounds by the high energy neutrinos are expected to be about 1\% of the total number of CENNS signal events. \par

%TAB==============
\begin{table}[t]
\begin{center}
	\begin{tabular}{  l  r  c  }
    \hline
                           & Uncertainty   		\\ \hline
    Neutrino flux          &       9\%           \\ 
    {\it Leff} of LAr      &       7.5\%      	\\ 
    High energy neutrinos  &$<$1\%            	\\ 
    Beam-induced neutrons  &$<$1\%            	\\ 
    Cosmogenic neutrons    &$<$1\%            	\\ 
    $^{39}$Ar and gammas   &$<$0.5\%          	\\ 
    Radiogenic backgrounds &$<$1\%         		\\ \hline
    \hline
    {\bf Total uncertainty}  &  {\bf 12\%}  \\ \hline
    \end{tabular}
    
    {\caption{ Systematic uncertainties of the event rate of CENNS experiment. The detector energy threshold is assumed to be E$_{th} \ge 25$\,keV$_{nr}$.}\label{tab:systematics}}
\end{center}
\end{table}
%TAB==============

\subsubsection{Uncertainties from beam-induced neutrons}
The neutron flux measurement by SciBath and results from a neutron shielding MC study indicate that the beam-induced neutrons can be substantially reduced with proper shielding design and could have a negligible impact on the CENNS event rates. However, due to the potential unknowns of these fast neutron shielding effects, and our current uncertainty in neutron sources and directions we assign a systematic uncertainty of beam-induced neutrons on the CENNS event rate at the 1\% level. \par

\subsubsection{Uncertainties from cosmogenic neutrons, gammas, radons and $^{39}$Ar}
The non-beam-related backgrounds can be significantly suppressed by the duty factor. Therefore the background requirement of the CENNS experiment is far less stringent than that of typical dark matter or other low background experiments. Cosmic-ray backgrounds can be further reduced by an active veto system in the water shielding, or it can also be significantly suppressed by 4\,m of passive concrete shielding (see FIG.~\ref{fig:NeutronShielding}). The expected systematic uncertainty of the cosmogenic neutrons events in the signal rate is less than 1\%. The gamma backgrounds are produced mostly by the decay chain of $^{238}$U, $^{232}$Th, and $^{40}$K in the PMT glass windows. These gamma backgrounds can also be suppressed by the duty factor, PSD and fiducial volume cuts. As shown in the FIG.~\ref{fig:evrate} the contribution of the gamma backgrounds in the signal region is negligible. The backgrounds from the radon daughters, especially $^{210}$Po can produce nuclear recoils in the signal region. The radon daughter backgrounds in the signal region is expected to be negligible after the pulse timing cut. Moreover, the steady-state backgrounds can be separately measured by the beam-off data in the energy region of interest and can be subtracted from the signal shape. Therefore, the systematic uncertainty due to radiogactive backgrounds is conservatively assumed to be less than 1\%.\par

%FIG==============
\begin{figure}[t]
\centering
\includegraphics[width=3.6in]{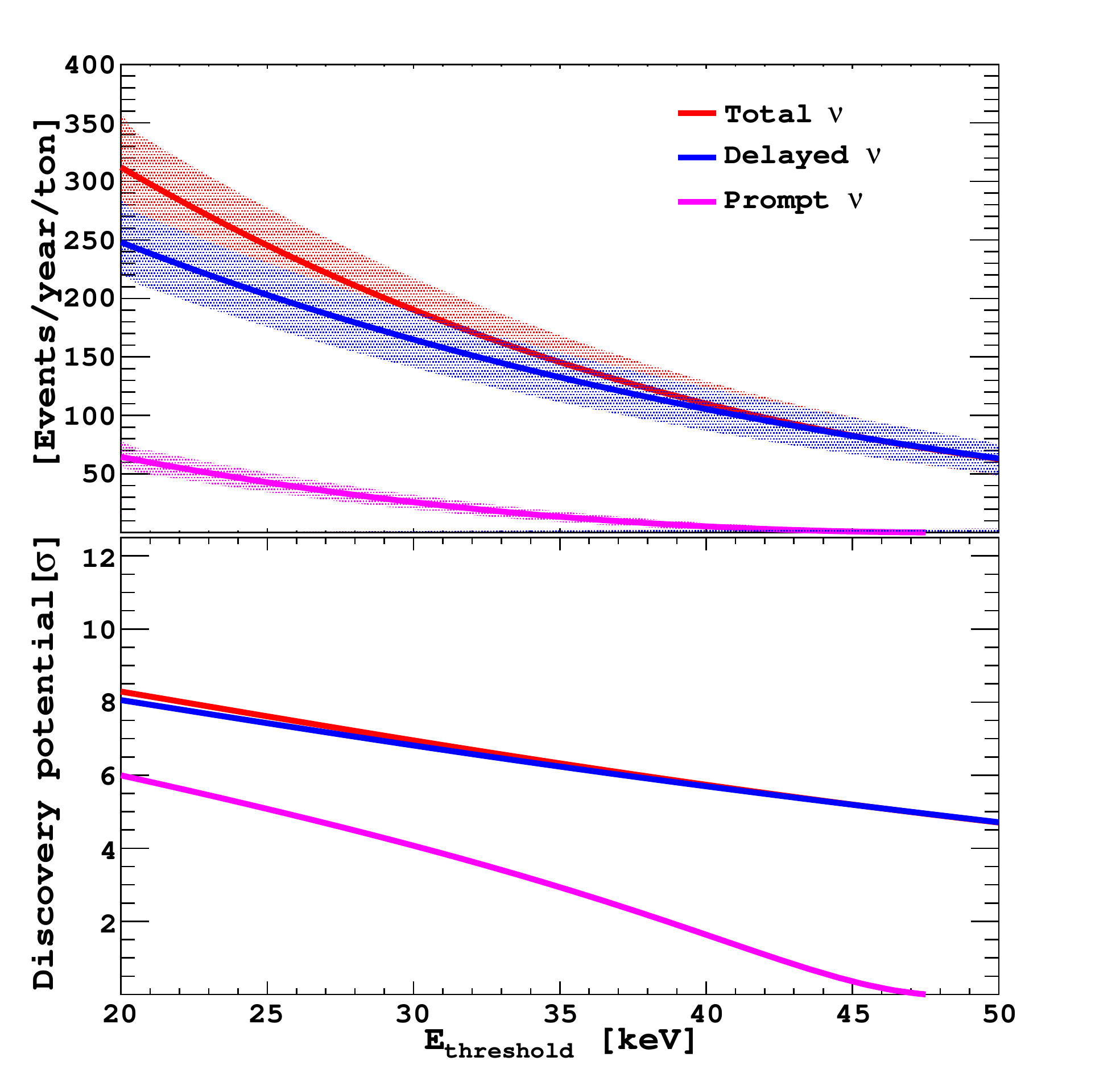}
{\caption{ CENNS discovery potential. The integrated signal event rates per ton detector after one-year operation as a function of detector energy threshold (top plot) and the discovery potential in $\sigma$ (bottom plot). A flat detection efficiency of 50\% over the energy range is assumed. The error bands on the top plot are 1 sigma quadratic-sum errors of statistical and systematic errors.}\label{fig:cennssensitivity}}
\end{figure}
%FIG==============

Table~\ref{tab:systematics} summarizes the systematic uncertainties. The total systematic uncertainty in event rate is expected to be 12\%. FIG.~\ref{fig:cennssensitivity} shows the discovery potential of the CENNS interaction as a function of detector energy threshold with 1\,ton-year exposure at 20\,m from the BNB target. A 7.5 sigma discovery of the CENNS is expected at the detector energy threshold of 25\,keV$_{nr}$. \par

%%%%%%%%%%%%%%%%%%%%%%%%%%%%%%%%%%%%%%%%%%%%%%%%%%%%%%%%%%%%%%%%%%%%%%%%%%%%%%%%%%%%%%%
\section{Summary}\label{summary}
%%%%%%%%%%%%%%%%%%%%%%%%%%%%%%%%%%%%%%%%%%%%%%%%%%%%%%%%%%%%%%%%%%%%%%%%%%%%%%%%%%%%%%%
We presented a new experimental method for measuring the Coherent Elastic Neutrino Nucleus Scattering (CENNS), utilizing low energy neutrinos emitted at the far-off-axis of a high energy neutrino beam. To determine the feasibility of this approach, we have made neutron background measurements at the Fermilab Booster Neutrino Beam (BNB). Our results indicate that this method can result in a successful experiment. With the BNB neutrino source, non-beam related backgrounds such as cosmic rays, internal and external radioactivity are substantially suppressed by the beam duty factor. The measured beam-induced neutron backgrounds can be safely reduced with proper shielding. We show that a one-ton fiducial mass single-phase liquid argon detector can make a 7.5 sigma discovery of CENNS at the detector energy threshold of 25\,keV$_{nr}$. Further development of a low energy neutrino source at Fermilab as part of programs like Project-X~\cite{Kronfeld:2013uoa} and nuSTORM~\cite{Adey:2013pio} will provide excellent resources for the future low energy neutrino physics experiments.\par

%%%%%%%%%%%%%%%%%%%%%%%%%%%%%%%%%%%%%%%%%%%%%%%%%%%%%%%%%%%%%%%%%%%%%%%%%%%%%%%%%%%%%%%
\section*{Acknowledgements}
%%%%%%%%%%%%%%%%%%%%%%%%%%%%%%%%%%%%%%%%%%%%%%%%%%%%%%%%%%%%%%%%%%%%%%%%%%%%%%%%%%%%%%%
The authors are grateful to R.~Davis, W.~Jaskierny, W.~Miner, K.~Taheri and J.~Volk for providing technical and engineering help of neutron flux measurement at the Fermilab Booster target building. We would also like to acknowledge H.~O.~Meyer for crucial work on the SciBath concept and detector construction. This work is supported by the Department of Energy, Fermilab University Research Association Visiting Scholar program, NSF-1068712, NSF-1306942 and NSF-1005233.

%{BIBLIOGRAPHY}
%%%%%%%%%%%%%%%%%%%%%%%%%%%%%%%%%%%%%%%%%%%%%%%%%%%%%%%%%%%%%%%%
\bibliographystyle{h-physrev3}
\bibliography{cennsreference}

\begin{thebibliography}{10}

\bibitem{Freedman:1973yd}
D.~Z. Freedman,
\newblock Phys. Rev. {\bf D9}, 1389 (1974).
%%CITATION = PHRVA,D9,1389;%%

\bibitem{Yoo:2011}
J.~Yoo,
\newblock Short-Baseline Neutrino Workshop (2011) ,
  https://indico.fnal.gov/conferenceDisplay.py?confId=4157.
%%CITATION = HEP-EX/0511001;%%

\bibitem{Wong:2005vg}
H.~T. Wong, H.-B. Li, J.~Li, Q.~Yue, and Z.-Y. Zhou,
\newblock J.Phys.Conf.Ser. {\bf 39}, 266 (2006), hep-ex/0511001.
%%CITATION = HEP-EX/0511001;%%

\bibitem{Barbeau:2007qi}
P.~Barbeau, J.~Collar, and O.~Tench,
\newblock JCAP {\bf 0709}, 009 (2007), nucl-ex/0701012.
%%CITATION = NUCL-EX/0701012;%%

\bibitem{Scholberg:2009ha}
K.~Scholberg {\em et~al.},
\newblock (2009), 0910.1989.
%%CITATION = ARXIV:0910.1989;%%

\bibitem{Akimov:2013yow}
CSI Collaboration, D.~Akimov {\em et~al.},
\newblock (2013), 1310.0125.
%%CITATION = ARXIV:1310.0125;%%

\bibitem{Jungman:1995df}
G.~Jungman, M.~Kamionkowski, and K.~Griest,
\newblock Phys. Rept. {\bf 267}, 195 (1996), hep-ph/9506380.
%%CITATION = HEP-PH/9506380;%%

\bibitem{Arrenberg:2008wy}
S.~Arrenberg, L.~Baudis, K.~Kong, K.~T. Matchev, and J.~Yoo,
\newblock Phys. Rev. {\bf D78}, 056002 (2008), 0805.4210.
%%CITATION = 0805.4210;%%

\bibitem{Billard:2013qya}
J.~Billard, L.~Strigari, and E.~Figueroa-Feliciano,
\newblock (2013), 1307.5458.
%%CITATION = ARXIV:1307.5458;%%

\bibitem{Cushman:2013zza}
P.~Cushman {\em et~al.},
\newblock (2013), 1310.8327.
%%CITATION = ARXIV:1310.8327;%%

\bibitem{Freedman:1977xn}
D.~Z. Freedman, D.~N. Schramm, and D.~L. Tubbs,
\newblock Ann.Rev.Nucl.Part.Sci. {\bf 27}, 167 (1977).
%%CITATION = ARNUA,27,167;%%

\bibitem{Nikkel:2006nh}
J.~A. Nikkel, R.~Hasty, W.~H. Lippincott, and D.~N. McKinsey,
\newblock Astropart. Phys. {\bf 29}, 161 (2008), astro-ph/0612108.
%%CITATION = ASTRO-PH/0612108;%%

\bibitem{Horowitz:2003cz}
C.~J. Horowitz, K.~Coakley, and D.~McKinsey,
\newblock Phys. Rev. D {\bf 68}, 023005 (2003).

\bibitem{Scholberg:2012id}
K.~Scholberg,
\newblock Ann.Rev.Nucl.Part.Sci. {\bf 62}, 81 (2012), 1205.6003.
%%CITATION = ARXIV:1205.6003;%%

\bibitem{Chakraborty:2013zua}
S.~Chakraborty, P.~Bhattacharjee, and K.~Kar,
\newblock (2013), 1309.4492.
%%CITATION = ARXIV:1309.4492;%%

\bibitem{Aguilar:2001ty}
LSND Collaboration, A.~Aguilar-Arevalo {\em et~al.},
\newblock Phys.Rev. {\bf D64}, 112007 (2001), hep-ex/0104049.
%%CITATION = HEP-EX/0104049;%%

\bibitem{Aguilar-Arevalo:2013pmq}
MiniBooNE Collaboration, A.~Aguilar-Arevalo {\em et~al.},
\newblock Phys.Rev.Lett. {\bf 110}, 161801 (2013), 1207.4809.
%%CITATION = ARXIV:1207.4809;%%

\bibitem{Garvey:2005pn}
G.~Garvey {\em et~al.},
\newblock Phys.Rev. {\bf D72}, 092001 (2005), hep-ph/0501013.
%%CITATION = HEP-PH/0501013;%%

\bibitem{Anderson:2012pn}
A.~Anderson {\em et~al.},
\newblock Phys.Rev. {\bf D86}, 013004 (2012), 1201.3805.
%%CITATION = ARXIV:1201.3805;%%

\bibitem{Vogel:1989iv}
P.~Vogel and J.~Engel,
\newblock Phys. Rev. {\bf D39}, 3378 (1989).
%%CITATION = PHRVA,D39,3378;%%

\bibitem{Bell:2006wi}
N.~F. Bell, M.~Gorchtein, M.~J. Ramsey-Musolf, P.~Vogel, and P.~Wang,
\newblock Phys.Lett. {\bf B642}, 377 (2006), hep-ph/0606248.
%%CITATION = HEP-PH/0606248;%%

\bibitem{Beda:2009kx}
A.~Beda {\em et~al.},
\newblock Phys.Part.Nucl.Lett. {\bf 7}, 406 (2010), 0906.1926.
%%CITATION = ARXIV:0906.1926;%%

\bibitem{PhysRevD.63.112001}
LSND Collaboration, L.~B. Auerbach {\em et~al.},
\newblock Phys. Rev. D {\bf 63}, 112001 (2001).

\bibitem{PhysRevLett.64.2856}
G.~G. Raffelt,
\newblock Phys. Rev. Lett. {\bf 64}, 2856 (1990).

\bibitem{PhysRevD.59.111901}
A.~Ayala, J.~C. D'Olivo, and M.~Torres,
\newblock Phys. Rev. D {\bf 59}, 111901 (1999).

\bibitem{Barranco:2005yy}
J.~Barranco, O.~G. Miranda, and T.~I. Rashba,
\newblock JHEP {\bf 12}, 021 (2005), hep-ph/0508299.
%%CITATION = HEP-PH/0508299;%%

\bibitem{Davidson:2003ha}
S.~Davidson, C.~Pena-Garay, N.~Rius, and A.~Santamaria,
\newblock JHEP {\bf 0303}, 011 (2003), hep-ph/0302093.

\bibitem{Scholberg:2005qs}
K.~Scholberg,
\newblock Phys. Rev. {\bf D73}, 033005 (2006), hep-ex/0511042.
%%CITATION = HEP-EX/0511042;%%

\bibitem{Barranco:2007tz}
J.~Barranco, O.~G. Miranda, and T.~I. Rashba,
\newblock Phys. Rev. {\bf D76}, 073008 (2007), hep-ph/0702175.
%%CITATION = HEP-PH/0702175;%%

\bibitem{Dorenbosch:1986tb}
CHARM Collaboration, J.~Dorenbosch {\em et~al.},
\newblock Phys.Lett. {\bf B180}, 303 (1986).
%%CITATION = PHLTA,B180,303;%%

\bibitem{Steiner2005325}
A.~Steiner, M.~Prakash, J.~Lattimer, and P.~Ellis,
\newblock Physics Reports {\bf 411}, 325  (2005).

\bibitem{Patton:2012jr}
K.~Patton, J.~Engel, G.~C. McLaughlin, and N.~Schunck,
\newblock Phys.Rev. {\bf C86}, 024612 (2012), 1207.0693.
%%CITATION = ARXIV:1207.0693;%%

\bibitem{Amanik:2009zz}
P.~Amanik and G.~McLaughlin,
\newblock J.Phys.G {\bf G36}, 015105 (2009).

\bibitem{Horowitz:2000xj}
C.~Horowitz and J.~Piekarewicz,
\newblock Phys.Rev.Lett. {\bf 86}, 5647 (2001), astro-ph/0010227.
%%CITATION = ASTRO-PH/0010227;%%

\bibitem{Elnimr:2013wfa}
OscSNS Collaboration, M.~Elnimr {\em et~al.},
\newblock (2013), 1307.7097.
%%CITATION = ARXIV:1307.7097;%%

\bibitem{Efremenko:2008an}
Y.~Efremenko and W.~Hix,
\newblock J.Phys.Conf.Ser. {\bf 173}, 012006 (2009), 0807.2801.
%%CITATION = ARXIV:0807.2801;%%

\bibitem{Bolozdynya:2012xv}
A.~Bolozdynya {\em et~al.},
\newblock (2012), 1211.5199.
%%CITATION = ARXIV:1211.5199;%%

\bibitem{bnbmap:1999}
Figure is taken from a MiniBooNE site plan, Drawing No. 6-7-52 CDR-2  (1999).

\bibitem{BNBDRAWING:2000}
R.~Stefanski {\em et~al.},
\newblock Fermilab internal (Project 6-7-55)  (2000).

\bibitem{AguilarArevalo:2008yp}
MiniBooNE Collaboration, A.~A. Aguilar-Arevalo {\em et~al.},
\newblock Phys. Rev. {\bf D79}, 072002 (2009), 0806.1449.
%%CITATION = 0806.1449;%%

\bibitem{BNBTDR:2001}
I.~Stancu,
\newblock Fermilab internal  (2001).

\bibitem{Tayloe:2006ct}
R.~Tayloe {\em et~al.},
\newblock Nucl.Instrum.Meth. {\bf A562}, 198 (2006).

\bibitem{Cooper:2011kx}
R.~Cooper {\em et~al.},
\newblock (2011), 1110.4432.
%%CITATION = ARXIV:1110.4432;%%

\bibitem{Verbinski}
V.~V. Verbinski {\em et~al.},
\newblock Nucl. Instr. Meth. {\bf 65}, 8 (1968).

\bibitem{Adams}
J.~M. Adams and G.~White,
\newblock Nucl. Instr. Meth. {\bf 156}, 495 (1978).

\bibitem{Cecil}
R.~A. Cecil, B.~D. Anderson, and R.~Madey,
\newblock Nucl. Instr. Meth. {\bf 161}, 439 (1979).

\bibitem{Bugel:2004yk}
FINeSSE Collaboration, L.~Bugel {\em et~al.},
\newblock (2004), hep-ex/0402007.
%%CITATION = HEP-EX/0402007;%%

\bibitem{Eljen}
{Eljen Technology, Sweetwater, TX 79556},
\newblock {www.eljentechnology.com/},
\newblock accessed Mar.~5, 2013.

\bibitem{StGobain}
{Saint-Gobain Crystals},
\newblock {http://www.detectors.saint-gobain.com/},
\newblock accessed Mar.~5, 2013.

\bibitem{Braizinha:2010zz}
B.~Braizinha, J.~Esterline, H.~Karwowski, and W.~Tornow,
\newblock Nucl. Instrum. Meth. {\bf A623}, 1046 (2010).
%%CITATION = NUIMA,A623,1046;%%

\bibitem{Mei:2005gm}
D.~Mei and A.~Hime,
\newblock Phys.Rev. {\bf D73}, 053004 (2006), astro-ph/0512125.
%%CITATION = ASTRO-PH/0512125;%%

\bibitem{Miyake:1973qk}
S.~Miyake,
\newblock {International Cosmic Ray Conference} {\bf 5}, 3638 (1973),
\newblock {Denver 1973}.
%%CITATION = INSPIRE-83114;%%

\bibitem{Gordon:2004}
M.~Gordon {\em et~al.},
\newblock IEEE.Trans.Nucl.Sci {\bf 51}, 3427 (2004).

\bibitem{PhysRevB.13.1649}
S.~Kubota {\em et~al.},
\newblock Phys. Rev. B {\bf 13}, 1649 (1976).

\bibitem{PhysRevB.27.5279}
A.~Hitachi {\em et~al.},
\newblock Phys. Rev. B {\bf 27}, 5279 (1983).

\bibitem{Doke:1988dp}
T.~Doke {\em et~al.},
\newblock Nucl. Instrum. Meth. {\bf A269}, 291 (1988).
%%CITATION = NUIMA,A269,291;%%

\bibitem{Doke2002}
T.~Doke {\em et~al.},
\newblock Jpn. J. Appl. Phys. {\bf 41}, 1538 (2002).

\bibitem{Boulay:2006mb}
M.~G. Boulay and A.~Hime,
\newblock Astropart. Phys. {\bf 25}, 179 (2006).
%%CITATION = APHYE,25,179;%%

\bibitem{Lippincott:2008ad}
W.~H. Lippincott {\em et~al.},
\newblock Phys. Rev. {\bf C78}, 035801 (2008), 0801.1531.
%%CITATION = 0801.1531;%%

\bibitem{Benetti:2007cd}
P.~Benetti {\em et~al.},
\newblock Astropart.Phys. {\bf 28}, 495 (2008), astro-ph/0701286.
%%CITATION = ASTRO-PH/0701286;%%

\bibitem{Hime2011}
A.~Hime,
\newblock Proceedings of the DPF-2011 Conference, Providence, RI.  (2011).

\bibitem{Alexander201344}
T.~Alexander {\em et~al.},
\newblock Astroparticle Physics {\bf 49}, 44  (2013).

\bibitem{Boulay:2009zw}
M.~Boulay {\em et~al.},
\newblock (2009), 0904.2930.
%%CITATION = ARXIV:0904.2930;%%

\bibitem{Catanesi:2007ab}
HARP Collaboration, M.~Catanesi {\em et~al.},
\newblock Eur.Phys.J. {\bf C52}, 29 (2007), hep-ex/0702024.
%%CITATION = HEP-EX/0702024;%%

\bibitem{DSchmitz2008}
D.~W. Schmitz,
\newblock PhD Thesis, Columbia University  (2008).

\bibitem{Regenfus:2006rt}
C.~Regenfus,
\newblock {IDM2006 Proceedings, World Scientific} , 325 (2006).
%%CITATION = INSPIRE-1191257;%%

\bibitem{Brunetti:2004cf}
WArP Collaboration, R.~Brunetti {\em et~al.},
\newblock New Astron. Rev. {\bf 49}, 265 (2005), astro-ph/0405342.
%%CITATION = ASTRO-PH/0405342;%%

\bibitem{Mei200812}
D.-M. Mei, Z.-B. Yin, L.~Stonehill, and A.~Hime,
\newblock Astroparticle Physics {\bf 30}, 12  (2008).

\bibitem{fluka1}
G.~Battistoni {\em et~al.},
\newblock AIP Conference Proceeding {\bf 896}, 31 (2007).

\bibitem{fluka2}
A.~Fass\`o, A.~Ferrari, J.~Ranft, and P.~Sala,
\newblock (2005).

\bibitem{fluka3}
G.~Battistoni, A.~Ferrari, M.~Lantz, P.~R. Sala, and G.~I. Smirnov,
\newblock (2009).

\bibitem{genie}
C.~Andreopoulos {\em et~al.},
\newblock Nucl. Instrum. Meth. {\bf A614}, 87 (2010), 0905.2517.
%%CITATION = 0905.2517;%%

\bibitem{Kronfeld:2013uoa}
A.~S. Kronfeld {\em et~al.},
\newblock (2013), 1306.5009.
%%CITATION = ARXIV:1306.5009;%%

\bibitem{Adey:2013pio}
nuSTORM Collaboration, D.~Adey {\em et~al.},
\newblock (2013), 1308.6822.
%%CITATION = ARXIV:1308.6822;%%

\end{thebibliography}

\end{document}